\pdfoutput=1
\documentclass[prd,aps,preprintnumbers,showpacs,10pt,onecolumn,nofootinbib]{revtex4}
\usepackage{amsfonts,amsmath,amssymb,bm,dsfont,eucal,slashed}
\usepackage[english]{babel}
\usepackage[latin1]{inputenc}
\usepackage[T1]{fontenc}
\usepackage{ae,aecompl}
\usepackage{graphicx,graphics,epsfig,epstopdf,subfigure,color,psfrag}
\usepackage{hyperref}
\newcommand{\bea}{\begin{eqnarray}}
\newcommand{\eea}{\end{eqnarray}}
\newcommand{\be}{\begin{equation}}
\newcommand{\ee}{\end{equation}}
\newcommand{\ba}{\begin{array}}
\newcommand{\ea}{\end{array}}
\newcommand{\nn}{\nonumber}

\newcommand{\bra}[1]{\left\langle #1 \right|}
\newcommand{\ket}[1]{\left| #1 \right\rangle}

\begin{document}

\title{Anomalous \boldmath $AV^*V$ vertex function \\ in the soft-wall holographic model of QCD}

\author{P. Colangelo}
\affiliation{Istituto Nazionale di Fisica Nucleare, Sezione di Bari, Italy}
\author{F. De Fazio}
\affiliation{Istituto Nazionale di Fisica Nucleare, Sezione di Bari, Italy}
\author{F. Giannuzzi}
\affiliation{Universit\`a degli Studi di Bari and Istituto Nazionale di Fisica Nucleare, Sezione di Bari, Italy}
\author{S. Nicotri}
\affiliation{Universit\`a degli Studi di Bari and Istituto Nazionale di Fisica Nucleare, Sezione di Bari, Italy}
\author{J. J.  Sanz-Cillero}
\affiliation{Istituto Nazionale di Fisica Nucleare, Sezione di Bari, Italy}


\begin{abstract}
We consider the vertex function of two vector and one axial-vector currents using the soft-wall holographic model of QCD with the Chern-Simons term. Two structure functions $w_L$ and $w_T$ describe such a vertex in the special case  in which one of the two vector currents corresponds to an on-shell  soft photon. We briefly review  the QCD results for these functions,  obtained from  triangular loop diagrams with quarks having  mass  $m_q=0$ or  $m_q \neq 0$, we compute $w_L$ and $w_T$ in the soft-wall model and compare the outcome to the QCD findings. We also calculate and discuss the two-point $\Pi_{VV}-\Pi_{AA}$ correlation function, together with a few  low-energy constants, which turn out to be close to the QCD results.  Finally, we comment on  a relation proposed by Son and Yamamoto between  $w_T$ and  $\Pi_{VV}-\Pi_{AA}$.
\end{abstract}


\maketitle

\section{Introduction}\label{sec:intro}
The anti-de Sitter/conformal field theory (AdS/CFT) correspondence conjecture \cite{Maldacena:1997re,Witten:1998qj,Gubser:1998} provides tools to access gauge theories at strong coupling. This remarkable result has inspired  the idea that
the  quantum chromodynamics   can be described using  methods rooted in the gauge/gravity duality principle, as first proposed in \cite{Witten2:1998}.
QCD is  different from the field theories for which the correspondence has been established; however,  being nearly conformal in the UV,  for massless quarks and neglecting the running of the strong coupling constant,  QCD can be considered as a candidate for a description based on gauge/gravity duality,  on condition  (at least)  that  a mechanism to break conformal invariance in the infrared region and  to generate confinement is supplied.
A strategy that can be pursued is the so-called ``bottom-up"  approach:  one starts from QCD and tries to construct a five-dimensional (5d) holographic dual theory encoding  as much as possible the
QCD properties,  namely hadron spectra, form factors, hadronic matrix elements.
The features of the dual theory are then scrutinized with the final purpose of selecting the best formulation which  (hopefully) can be used to compute  properties of QCD  not accessible to other analytical or  numerical approaches.
Under the name of AdS/QCD a number of extra-dimensional models are collected,  set up  with the aim of  reproducing  the largest number of known  QCD aspects
\cite{altri,Erlich:2005qh,Da Rold:2005zs,Karch:2006pv,Colangelo:2008us} \footnote{We do not discuss here the so-called ``top-down" AdS/QCD approach, which has been  reviewed, e.g., in \cite{top-down}.}.

 An important point to investigate in the holographic approaches is related to the chiral anomaly.  It is known that  the longitudinal part of massless fermion anomalous triangle diagrams is fixed by the chiral anomaly,
which  produces, for example,  the successful expression of the $\pi^0 \to \gamma \gamma$ decay amplitude
 \cite{Adler:1969gk,Bell:1969ts, Adler:1969er}. For the transverse part of these triangle diagrams,  results have been obtained for current-current correlators in an infinitesimally weak electromagnetic field,  and such results concern the existence and the expression of both
perturbative and nonperturbative effects.  In particular, it has been found that, for massless quarks, radiative $\alpha_s$ corrections  are  absent also in the  transverse part of triangle diagrams, and that the  nonperturbative corrections show up  in this part
 at precise orders in the operator product  expansion (OPE)  \cite{Vainshtein:2002nv,Knecht:2003xy}. Other corrections  appear, both in the longitudinal and in the transverse part, if the quark masses do not vanish  \cite{VVA-Vainshtein,Melnikov:2006qb}.

The  investigation of  this sector of QCD using  holography  could permit to  assess  the degree of reliability of the gauge/gravity duality  approach  to the quantum chromodynamics, and indeed a few studies have been devoted to this  and other closely  related topics  in various  dual  models
\cite{Grigoryan:2008up,Gorsky:2009ma,Brodsky:2011xx}. In particular,  it has been suggested,  using a holographic model of QCD  in which the chiral symmetry is broken,  as  in \cite{Hirn:2005nr}, by boundary conditions for the vector and axial-vector fields,  that a relation should connect  the transverse part of the anomalous quark triangle diagrams and the two-point left-right current correlation function  \cite{Son:2010vc}.   However,  it has been claimed that such a relation is not obeyed in QCD~\cite{Knecht:2011wh}.

Motivated by the discussion, we consider the issue of the quark triangle  diagrams in a holographic model in which chiral symmetry breaking is realized by the presence of a
scalar field, as in \cite{Erlich:2005qh,Da Rold:2005zs},  and confinement is provided by a background dilaton field which ensures linear Regge trajectories for light hadrons, the so-called  soft-wall dual model of QCD \cite{Karch:2006pv}. Our  aim is to  compute  the longitudinal and transverse parts of the anomalous quark triangle diagrams  and establish which QCD features are reproduced in the  holographic  framework,  and whether relations exist between the transverse part and the left-right current correlation function,  as proposed in    \cite{Son:2010vc}. This also allows us to investigate in details aspects of the chiral symmetry breaking in the
soft-wall model.

We start our study by reviewing in Sec.\ref{sec:qcd}  the properties, in QCD, of the longitudinal and transverse part of anomalous triangle diagrams for  zero and nonvanishing quark mass. In Sec.\ref{sw}
we formulate the holographic soft-wall model with a Chern-Simons term, and in Sect.\ref{numerics} we determine the longitudinal and transverse structure functions $w_L$ and $w_T$ for various possibilities of the chiral symmetry breaking quantities, the quark mass and the quark condensate, collecting in the appendices  several  computational details.
The relations to two-point correlation functions, together with the properties of such correlation  functions,  are discussed in Secs.\ref{sec:PiAAVV} and \ref{sec:PiLR},  with a determination of a few low-energy constants.
 In Sec.\ref{concl} there are our conclusions.

\section{$AV^*V$ vertex function in QCD}\label{sec:qcd}
Let us consider  the vertex function involving two  vector currents $J_\mu=\bar q V \gamma_\mu q$
and  an axial-vector current  $J^5_\nu=\bar q A \gamma_\nu \gamma_5 q$, with quark fields $q^i_f$ carrying a color $(i)$ and a flavour $(f)$ index,
 and $V$ and $A$ diagonal matrices acting on the flavour indices.  In particular, we consider the case where one of the two vectors
corresponds to  a real and soft photon, i.e. with squared four-momentum $k^2=0$ and momentum $k\simeq 0$.
An example  of such a kind of functions is the  $Z^0\gamma^* \gamma$ vertex,  described by two electromagnetic  currents $J_\mu=\sum_f Q_f \bar q_f \gamma_\mu q_f$ with $Q_f$ the electric charges, and $J_\nu^5$  the axial current $J_\nu^5= \sum_f 2 I^3_{f} \, \bar q_f \gamma_\mu \gamma_5 q_f$
with $I^3_f$ the third component of the weak isospin, and in this case the sum involves the  quarks and also the leptons.
The triangle graph corresponding to the vertex produces the anomaly of the $Z^0$ axial current, which vanishes in the standard $SU(3)_c \times SU(2)_L\times U(1)_Y$ model  provided that the contributions of  all the fermions
(quarks and leptons) in a given generation are added up.

We define the two-point correlation function of $J_\mu$ and $J^5_\nu$ in an external electromagnetic field
\be
T_{\mu \nu}(q,k)=i \,\, \int d^4x \, e^{i\,q \cdot x} \bra{0} T[J_\mu(x)J_\nu^5(0)] \ket{\gamma(k,\epsilon)} \,\,\, .
\label{twopoint}
\ee
It  can be related to the three-point vacuum correlation function
\be
  T_{\mu \nu \sigma} (q,k)=i^2  \,\, \int d^4x \, d^4y\, e^{i\,q \cdot x-i \, k \cdot y} \bra{0} T[J_\mu(x)J_\nu^5(0)J^{em}_\sigma(y)] \ket{0}\,\,
\label{threepoint}
\ee
where  $J^{em}_\sigma$  is the electromagnetic current, since
\be
 T_{\mu \nu}(q,k)= e \,\epsilon^{ \sigma} \,T_{\mu \nu \sigma}(q,k) \,\,\, ,
\label{threepoint-relation}
\ee
with $\epsilon^\sigma(k)$  the photon polarization vector and $e$ the electric charge unit.

For soft photon momentum $k\to 0$ one can express  $T_{\mu \nu}(q,k)$  keeping only linear terms  in $k$ and neglecting  quadratic and higher order powers of the momentum.  In this kinematical condition, accounting for the conservation of the vector current $J_\mu$,   the amplitude $ T_{\mu \nu}$ can be decomposed
in terms of two structure functions $w_L(q^2)$  and $w_T(q^2)$:
\be
T_{\mu \nu}(q,k)=-{i \, \over 4 \pi^2} {\rm Tr}\left[Q V A\right] \left\{ w_T(q^2)(-q^2 {\tilde f}_{\mu \nu}+q_\mu q^\lambda {\tilde f}_{\lambda \nu} -q_\nu q^\lambda {\tilde f}_{\lambda \mu})+ w_L(q^2) q_\nu q^\lambda {\tilde f}_{\lambda \mu} \right\}\,\,, \label{decomposition}
\ee
where  $Q$ is the electric charge matrix and  ${\tilde f}_{\mu \nu}=\displaystyle{1 \over 2} \epsilon_{\mu \nu \alpha \beta} f^{\alpha \beta}$ is the  dual field  of the photon field strength $f^{\alpha \beta}=k^\alpha \epsilon^\beta -k^\beta \epsilon^\alpha$. The first term in the decomposition (\ref{decomposition}) is transversal  with respect to the axial current index, the second one is longitudinal.

We briefly recall what is known in QCD about the two invariant functions $w_{L}(q^2)$ and $w_{T}(q^2)$;   in the next Sections we shall compute  these quantities in the AdS/QCD soft-wall  model,  aiming  at   understanding which  QCD properties are reproduced in that holographic approach.

In the case in which the triangle loop corresponding to (\ref{twopoint}) and (\ref{threepoint})  takes contribution from a single  quark of mass $m$ belonging to the fundamental representation of the color gauge group $SU(N_c)$,  defining $Q^2=-q^2$, the one-loop result for $T_{\mu \nu}$ gives \cite{Adler:1969gk}
\be
w_L(Q^2)=2 \, w_T(Q^2) = {2 N_c \over Q^2} \left[1 +{2 m^2 \over Q^2}\ln{m^2 \over Q^2}+{\cal O} \left({m^4 \over Q^4} \right) \right] \,\,. \label{wLT-1loop} \ee
In principle, such a result could be modified by  perturbative and nonperturbative corrections.
Actually, a nonrenormalization theorem for the anomaly protects $w_L$ from receiving perturbative corrections \cite{Adler:1969er}. As for $w_T$, in \cite{Vainshtein:2002nv} it has been demonstrated that for the special kinematic condition considered here,  in which one of the photons is on shell and soft $(k \to 0$),  and for $Q^2 \gg m^2$,  the perturbative corrections to $w_T$ also vanish to all orders. This implies that
in the chiral limit $m=0$ the $\alpha_s$ corrections are both absent in $w_L$ and $ w_T$; hence,
\be
w_L(Q^2) = {2 N_c \over Q^2}\,\,
\label{basic-rel-0}\ee
and, discarding nonperturbative corrections, the relation holds:
\be
w_L(Q^2)=2 \, w_T(Q^2)\,\, .
\label{basic-rel}\ee

Now we turn to  the nonperturbative corrections in the case of  light quarks.
In the chiral limit  $m=0$ such corrections to $w_L$ are also absent,  a consequence of the fact that the behavior $\displaystyle w_L \propto {1 \over Q^2}$ reflects the contribution of the pion pole to the longitudinal part of $T_{\mu\nu}$, and the pole   is located in this case at $Q^2=0$.
On the other hand, $w_T$ receives nonperturbative corrections which start from $\displaystyle{{\cal O}\left({1 \over Q^6} \right)}$.

To understand the case $m \neq 0$,  we  consider  the nonperturbative corrections in the  framework of the OPE.
At large Euclidean $Q^2$ we define the  expansion of the operator ${\hat T}_{\mu \nu}$
\be
{\hat T}_{\mu \nu}=i \,\, \int d^4x \, e^{i\,q \cdot x}  \, T[J_\mu(x)J_\nu^5(0)] =\sum_i c^i_{\mu \nu \alpha_1 \alpha_2 \dots \alpha_i}(q) \, O_i^{\alpha_1 \alpha_2 \dots \alpha_i}
\label{T-hat}
\ee
in terms of   local operators  $O_i$ and of perturbatively computable coefficients $c^i$. The dimension of the operators $ O_i$ matches the dependence of the coefficients $c^i$ on the inverse powers of $Q^2$.
From the expansion (\ref{T-hat}) it  follows that
\be
T_{\mu \nu}(q,k)=\bra{0} {\hat T}_{\mu \nu} \ket{\gamma(k,\epsilon)}= \sum_i c^i_{\mu \nu \alpha_1 \alpha_2 \dots \alpha_i}(q) \bra{0} O_i^{\alpha_1 \alpha_2 \dots \alpha_i} \ket{\gamma(k,\epsilon)}\,\,.
\label{T-ope}
\ee
Keeping only  linear terms in the photon momentum $k$, the structure of  the OPE  for ${\hat T}_{\mu \nu}$ is
\be
{\hat T}_{\mu \nu}=\sum_i \left\{ c^i_T(q^2)(-q^2  O^i_{\mu \nu}+q_\mu q^\lambda  O^i_{\lambda \nu} -q_\nu q^\lambda  O^i_{\lambda \mu})+ c^i_L(q^2) q_\nu q^\lambda O^i_{\lambda \mu} \right\}\,\,,  \label{T-hat-ope}
\ee
so that, parameterizing the photon-vacuum matrix elements of the local operators $O_i$  as
\be
\bra{0} O_i^{\alpha \beta} \ket{\gamma(k,\epsilon)}=-{i \, e  \over 4 \pi^2} \kappa_i {\tilde f}^{\alpha \beta} 	 \,\,\,\, , \label{matrixelements}\ee
one has an expression for the functions $w_L$ and $w_T$ in terms of the coefficients $c^i$ and of the parameters $\kappa_i$,
\be
w_{L,T}(Q^2)=\sum_i c^i_{L,T}(Q^2) \, \kappa_i \,\,. \label{w-da-ci}
\ee
The leading (lowest dimensional) operator in the OPE has dimension  $D=2$ and  involves  the dual of the field strength tensor $F^{\alpha \beta}=\partial^\alpha A^\beta-\partial^\beta A^\alpha$, with $A$  the photon field:
\be
 O_{\alpha \beta}^{(D=2)}={e \over 4 \pi^2} {\tilde F}_{\alpha \beta}\,\,.
\ee
From the  relation $\bra{0} F_{\alpha \beta} \ket{\gamma(k,\epsilon)}=-i f_{\alpha \beta}$ and using the definition in (\ref{matrixelements})  one obtains $\kappa_{(D=2)}=1$.

The next contribution to the OPE comes from the  operator of dimension $D=3$
\be
O_{\alpha \beta}^{(D=3)}=-i {\bar q}\sigma_{\alpha \beta} \gamma_5 q\,\,
\ee
with coefficient   $\displaystyle c_{L,T}^{(D=3)}={4 m \over Q^4}$. From the relation $\sigma_{\alpha \beta} \gamma_5={i \over 2} \epsilon_{\alpha \beta \rho \tau}\sigma^{\rho \tau}$ and defining
\bea
\bra{0}{\bar q} \sigma^{\rho \tau} q \ket{\gamma(k,\epsilon)}&=&-{i \, e  \over 4 \pi^2} \kappa_{(D=3)} { f}^{\rho \tau} \nonumber \\
\kappa_{(D=3)}&=&-4 \pi^2 \langle {\bar q}q \rangle \chi
\eea
one obtains
\be
w_L^{(D=3)}(Q^2)=2w_T^{(D=3)}(Q^2)={4m \over Q^4} (-4 \pi^2) \langle {\bar q}q \rangle \chi
\label{wLT-D=3}
\ee
where $\langle \bar{q}q\rangle$ denotes the vacuum quark condensate and
we have introduced   the  so-called  magnetic susceptibility $\chi$ of the  quark condensate.
Therefore, at this order a  relation holds for $w_L$ and $w_T$:
\be
w_L(Q^2)=2 \, w_T(Q^2) = {2 N_c \over Q^2} \left[1 +{2 m^2 \over Q^2}\ln{m^2 \over Q^2} - {8 \pi^2 m \langle {\bar q}q \rangle \chi  \over N_c Q^2}+{\cal O} \left({m^4 \over Q^4} \right)\right] \,\, \label{wLT-2}
\ee
at large $Q^2$ (with ${\cal O}(\alpha_s)$ corrections computed in \cite{Melnikov:2006qb}). 
As for higher order terms, the dimension $D=4$ operators can be reduced to the $D=3$ ones  using the quark equation of motion, while both $D=5$ and $D=6$ terms contribute to ${\cal O} \left( {1 \over Q^6} \right)$ order.
Remarkably, the contribution of the dimension $D=6$ operators does not vanish in the chiral limit and is responsible of the difference between $w_L$ and $2\, w_T$.
Indeed, for  $m_q=0$,   $w_L$ remains $\displaystyle w_L(Q^2)=\frac{2 N_c}{Q^2}$,  while  $w_T$, including  the leading nonperturbative correction,  reads~\cite{Knecht:2002hr,VVA-Vainshtein}:
\begin{eqnarray}
w_T(Q^2)  &=& \frac{N_c}{Q^2}\,\,
+\,\, \frac{128 \pi^3 \alpha_s \,\chi\,\langle\bar{q}q\rangle^2}{9\, Q^6}\, \,\,\,
+\,\,\,\, {\CMcal O}\left(\frac{1}{Q^8}\right)\, .
\label{eq.wT-OPE-mq0}
\end{eqnarray}
The susceptibility of the chiral condensate $\chi$
 arises  here after assuming factorization of the matrix element of  four-quark operators
in the electromagnetic external field  $F^{\alpha\beta}$.
In principle,  there might  be other ${\CMcal O}(1/Q^6)$ contributions in the OPE from  operators like
$\widetilde{F}^{\alpha\beta} G^a_{\mu\nu}G_a^{\mu\nu}$, with $G^a_{\mu\nu}$ the gluon field strength;  however, they appear at one-loop  with small coefficients,  while the $1/Q^6$ term in~(\ref{eq.wT-OPE-mq0})
comes from  tree-level diagrams.

In the next sections we discuss the determination of the functions $w_{L}(Q^2)$ and $w_{T}(Q^2)$ in the soft-wall model,  to assess the  extent to which these QCD results are reproduced.

\section{The soft-wall AdS/QCD model with the Chern-Simons term}\label{sw}

As in other holographic approaches, the AdS/QCD soft-wall model \cite{Karch:2006pv} is defined in a five-dimensional AdS space with line element
\be
ds^2=g_{MN}dx^M dx^N={R^2 \over z^2}(\eta_{\mu \nu} dx^\mu dx^\nu-dz^2)\,\,.
\label{metric}
\ee
The coordinate  indices $M,N$ are $M,N=0,1,2,3,5$,  $\eta_{\mu \nu}=diag(+1,-1,-1,-1)$ and $R$  is the AdS curvature radius (set to unity from now on).
In the model,  the fifth coordinate $z$ runs in the range $\epsilon \le z < + \infty$,
with $\epsilon \to 0^+$, and a background dilatonlike field  is introduced
\be
\Phi(z)=(cz)^2 \label{dilaton}\,\,,\ee
the form of which is chosen to obtain  linear Regge trajectories for light vector mesons; $c$ is a dimensionful parameter setting a scale for QCD quantities, and its numerical value, obtained from the
spectrum of the light vector mesons, is $c={M_\rho \over 2}$. The model describes light  vector,  axial-vector  and pseudoscalar mesons, with a mechanism of chiral symmetry breaking  related to
the presence of a scalar field;  the model  has been  extended  to include  the sector of light scalar mesons \cite{Colangelo:2008us}.

As in \cite{Erlich:2005qh,Da Rold:2005zs,Karch:2006pv}, we introduce
the left and right gauge fields ${\cal A}_{L\mu}^a$ and ${\cal A}_{R\mu}^a$ which are
dual to the $SU(N_f)_L$ and $SU(N_f)_R$ flavour currents,  ${\bar q}_L \gamma^\mu T^a q_L$ and ${\bar q}_R \gamma^\mu T^a q_R$, with $T^a$  the generators of $SU(N_f)$.
 Moreover, we enlarge the gauge group to $U(N_f)_L \times U(N_f)_R$ to describe the dual of the electromagnetic  current which contains both isovector and isoscalar components.

 We  introduce a scalar field $X$ which is the dual to the quark bifundamental field ${\bar q}_R^\alpha q_L^\beta$:
\be
X=X_0 e^{2i\pi}
\label{X}
\ee
and contains the background field $X_0=\displaystyle{v(z) \over 2}$ and the chiral field $\pi(x,z)$. $X_0$ only depends on $z$ and  incorporates the chiral symmetry breaking behavior.
A scalar  field $S(x,z)$ could also be included to describe light scalar mesons  by the substitution $X_0 e^{2i\pi} \to (X_0 + S) e^{2i\pi} $  \cite{Colangelo:2008us}.
It is represented by $S(x,z)=S^A(x,z) \,T^A$,
with the indices $A=0,a$ and $a=1,\dots N_f^2-1$.    The matrix $T^0=\displaystyle{1 \over \sqrt{2N_f}}$,
together with   the $SU(N_f)$ generators  $T^a$, satisfies the normalization condition
\be
Tr(T^A \, T^B)={\delta^{AB} \over 2} \label{traccia-Ta}
\,\,.
\ee
The five-dimensional Yang-Mills  action describing the  fields  ${\cal A}_{L,R}^M$, as well as  the $X$ field,  is
\be
 S_{YM}={1 \over k_{YM}} \int d^5x \sqrt{g}e^{-\Phi} Tr\left\{ |DX|^2-m_5^2 |X|^2 -{1 \over 4 g_5^2} (F_L^2+F_R^2) \right\} \,\,\, ,
 \label{action-0}
 \ee
 with $F_{L,R}^{MN}=F_{L,R}^{MNa}T^a=\partial^M {\cal A}^N_{L,R}-\partial^N {\cal A}^M_{L,R}-i \left[ {\cal A}^M_{L,R}, {\cal A}^N_{L,R} \right]$.
 $g$ is the determinant of the metric tensor $g_{MN}$, $\Phi(z)$ is the dilaton  in (\ref{dilaton}),  and $k_{YM}$ is a parameter included to provide canonical $4d$ dimensions for the fields.
 The 5d mass of the field $X$  is fixed to $m_5^2=-3$ according to the AdS/CFT correspondence dictionary. The covariant derivative acting on $X$  is defined as
 \be
 D^M X=\partial^M X-i{\cal A}_L^MX+iX {\cal A}^M_R\,\,\,,
 \ee
hence for $X=0$ the left and right sectors in (\ref{action-0}) are decoupled.
 We combine the  gauge fields ${\cal A}_{L,R}^M$ into  a vector field $V^M=\displaystyle{{\cal A}_L^M +{\cal A}_R^M \over 2}$ and an axial-vector field  $A^M=\displaystyle{{\cal A}_L^M -{\cal A}_R^M \over 2}$,
 so that the 5d action for  the  fields ${V,A}$ and $X$ is
 \be
 S_{YM}={1 \over k_{YM}} \int d^5x \sqrt{g}e^{-\Phi} Tr\left\{ |DX|^2-m_5^2 |X|^2 -{1 \over 2 g_5^2} (F_V^2+F_A^2) \right\} \,\,\, .
 \label{action}
 \ee
 The covariant derivative  is  now  defined as
 \be
 D^MX=\partial^M X-i[V^M,X]-i\{A^M,X\} \,\,\,
 \ee
 and the field strengths $F_{V,A}^{MN}$ are
 \bea
 F_V^{MN}&=&\partial^M V^N -\partial^N V^M -i[V^M,V^N]-i[A^M,A^N] \nonumber \\
 F_A^{MN}&=&\partial^M A^N -\partial^N A^M -i[V^M,A^N]-i[A^M,V^N] \,\,\, .
 \eea
 Matching the two-point correlation function of the vector field $V$, and the two-point  correlation function of the  scalar field $S$, with the corresponding leading order perturbative QCD results allows to fix the constants
 $k_{YM}$ and $g_5^2$ in the Yang-Mlls action:  $\displaystyle k_{YM}={16 \pi^2 \over N_c}$ and
 $\displaystyle g_5^2={3 \over 4}$ \cite{Erlich:2005qh,Colangelo:2008us}.

 The modification to the approach in \cite{Erlich:2005qh,Da Rold:2005zs,Karch:2006pv},  required to compute the functions $w_L$ and $w_T$, consists in adding  to $S_{YM}$  a Chern-Simons  contribution,  discussed in
 \cite{Witten:1998qj} and  considered  in holographic models in
 \cite{Hill:2006wu} and   \cite{Grigoryan:2008up,Gorsky:2009ma,Brodsky:2011xx,Son:2010vc,Domokos:2007kt,Gorsky:2010xu}. 
  This contribution is given by  the difference $S_{CS}({\cal A}_{L})-S_{CS}({\cal A}_{R})$, where
 \be
S_{CS}({\cal A})= k_{CS}\int d^5x \,\, Tr \left[ {\cal A} F^2-{i\over 2} {\cal A}^3 F - {1\over 10} {\cal A}^5 \right]. \label{chernsimons}
 \ee
Actually,  the  terms in  the Chern-Simons action $S_{CS}$ proportional to higher odd powers of ${\cal A}_{L,R}$ do not contribute to the correlation function $AV^*V$ of  interest here, therefore we do no consider them anymore, and only keep in (\ref{chernsimons}) the terms
 $\displaystyle Tr \left[ {\cal A}_{L,R} F^2_{L,R}\right]=\epsilon_{ABCDE}\, Tr \left[{\cal A}^A_{L,R} F_{L,R}^{BC} F_{L,R}^{DE} \right]$, with  $A,  \dots, E$  indices of the $5d$ coordinates.
 Moreover, since  the Chern-Simons actions  depend explicitly on the gauge fields $\cal A$ and are invariant only up to a boundary term, we include  a boundary term to make explicit the invariance under a
 vector gauge transformation, obtaining:
 \be
 S_{CS+b} = 3\,  k_{CS} \,\,  \epsilon_{ABCDE} \,  \int d^5x \,\,Tr \left[{A}^A \left\{F_{V}^{BC}, F_{V}^{DE} \right \} \right]  \,\,\, . \label{S-CS-boundary}
 \ee
The constant $k_{CS}$ will be fixed below. \footnote{In some top-down models of holographic QCD  the Chern-Simons action also contains  a coupling with the scalar tachyon $X$, as derived by brane actions  \cite{Kennedy:1999nn,Casero:2007ae}.}
In the AdS/QCD soft-wall model the starting point is then the  effective action
\be
S_{5d}^{eff}=S_{YM}+S_{CS+b}\,\,. \label{effective-action}
\ee

In order to compute correlation functions of  vector and  axial-vector currents,  we exploit the basic relation of the AdS/QCD correspondence, i.e. the duality relation between the QCD generating functional relative to a given operator $O(x)$ and the effective 5d action. The duality  holds provided that the source of $O(x)$ coincides with the $z=0$ boundary value $f_0(x)=f(x,0)$ of the dual field $f(x,z)$ in the 5d action:
\begin{equation}\label{generating}
  \biggl\langle e^{i\int d^4 x\;0(x )\,f_0(x)}\biggr\rangle_{QCD}=
  e^{iS_{5d}^{eff}[f(x,z)]}\,\,.
\end{equation}

Let us define ${\tilde G}^a_\mu(q,z)$ as the Fourier transform with respect to the 4d coordinates $x^\mu$ of a generic gauge field $G^a(x,z)=V^a(x,z)$ and $A^a(x,z)$
($a$ flavour index).  The bulk-to-boundary propagator $G(q,z)$ can analogously be defined in the Fourier space: ${\tilde G}^a_\mu(q,z)=G(q,z)G_{\mu 0}^a(q)$,  where $G_{\mu0}^a(q)$ is the source field.
Furthermore, we decompose each vector and axial-vector field of momentum $q$ using  two projection tensors,
\bea
P_{\mu \nu}^\perp &=& \eta_{\mu \nu}-{q_\mu q_\nu \over q^2} \nonumber
\\
P_{\mu \nu}^\parallel &=& {q_\mu q_\nu \over q^2} \,\,, \label{projectors}
\eea
so that the  vector and axial-vector bulk-to boundary propagators  can be written in terms of  the transverse and longitudinal parts:
\bea
{\tilde V}^a_\mu(q,z)&=&V_\perp(q,z)P_{\mu \nu}^\perp V_{ 0}^{a\nu }(q)\nonumber \\
{\tilde A}^a_\mu(q,z)&=&A_\perp(q,z)P_{\mu \nu}^\perp A_{0}^{a\nu}(q)+A_\parallel(q,z)P_{\mu \nu}^\parallel A_{ 0}^{a\nu}(q) \,\,,\label{perp-par} \eea
with boundary conditions $V_\perp(q,0)=1$ and $A_\perp(q,0)=A_\parallel(q,0)=1$.  We discuss below the behavior at   $z \to \infty$.
In (\ref{perp-par}) we have taken into account that the (conserved) vector field is transverse.

Writing the longitudinal component of ${\tilde A}$ as ${\tilde A}^{a \parallel}_\mu(q,z)=A_\parallel(q,z)P_{\mu \nu}^\parallel A_{\nu 0}^a(q)=i\,q_\mu {\tilde \phi}^a$,  from the effective 5d action
(\ref{effective-action}) we may work out a set of linearized equations of motion, obtained in the axial gauge $V_z=A_z=0$:

\bea
&&\partial_y \left({e^{-y^2} \over y}  \,\, \partial_y V_\perp \right)-{\tilde Q}^2 {e^{-y^2} \over y}V_\perp=0 \label{EOM-V} \\
&&\partial_y \left({e^{-y^2} \over y}  \,\, \partial_y A_\perp  \right)-{\tilde Q}^2 {e^{-y^2} \over y}A_\perp -{g_5^2 v^2(y) e^{-y^2} \over y^3}A_\perp=0 \label{EOM-A} \\
&&\partial_y \left({e^{-y^2} \over y} \,\, \partial_y  {\tilde \phi}^a  \right)+{g_5^2 v^2(y) e^{-y^2} \over y^3}  ({\tilde \pi}^a-{\tilde \phi}^a)=0 \label{EOM-phi-pi-1} \\
&&{\tilde Q}^2 (\partial_y  {\tilde \phi}^a) +{g_5^2 v^2(y) \over y^2}\partial_y {\tilde \pi}^a=0 \label{EOM-phi-pi-2} \,\,\, .\eea
We have defined the dimensionless quantities: $y=cz$ and ${\tilde Q}^2=\displaystyle{Q^2 \over c^2}$, with  $Q^2=-q^2$ ($Q^2>0$ represent  the  Euclidean momenta).
We  also adopt the  notation $V=V_\perp$ and $A=A_\perp$.  Using the relation
\begin{equation}
 {\tilde \phi}^a(q,y)=-i{q^\mu \over q^2}A_\parallel(q,y) P_{\mu \nu}^\parallel A_{\nu 0}^a(q) \,\,\, ,  \label{phitilde}
\end{equation}
and writing $\displaystyle {\tilde \pi}^a(q,y)=-i{q^\mu \over q^2}\pi(q,y) A_{\mu 0}^a(q)$, we find that  $\pi(q,y)$ and $A_\parallel(q,y)$ obey the same equations (\ref{EOM-phi-pi-1}) and (\ref{EOM-phi-pi-2})
as $\tilde \pi^a$ and $\tilde \phi^a$.

From the action (\ref{action}) an equation can also be derived for the field $X_0 = \frac{1}{2} \, v$:
\be
\partial_y \left({e^{-y^2} \over y^3}  \,\, \partial_y v(y)  \right)+{3 e^{-y^2} \over y^5} v(y)=0  \label{EOM-condensate}
\ee
the general solution of which is a  combination of the Tricomi confluent hypergeometric function $\displaystyle U\left({1\over 2}, 0, y^2\right)$ and of the Kummer confluent hypergeometric function $\displaystyle _1F_1\left({3\over 2}, 2, y^2\right)$. Imposing regularity of the solution for $y \to +\infty$,
the latter  must be discarded,  and  $v(y)$ reads
\be
 v(y)\sim \Gamma \left({3 \over 2}\right) \, y \, U\left({1\over 2}, 0, y^2\right) \,\,\,. \label{v-exact}
 \ee
 In the expansion of this function for $y \to 0$:   $v(y)\to C_1 y + C_2 y^3$, the two chiral symmetry breaking parameters can be identified on the basis of the holographic dictionary \cite{Erlich:2005qh}:  the quark mass,
 which breaks explicitly the chiral  symmetry, enters in the coefficient $C_1$ of $y$, and the quark condensate, the spontaneous chiral symmetry breaking parameter,  enters in the coefficient $C_2$ of $y^3$:
 \be
 m_q \propto C_1 \,\,\,\,\,\,\,\,\,\,  \sigma \propto \langle {\bar q} q\rangle \propto C_2 \,\,\, .
 \ee
 However, in the expansion of the solution $v(y)$ in (\ref{v-exact})
 the coefficients $C_1$ and $C_2$ are related, and this would imply a proportionality relation between the quark mass $m_q$ and the quark condensate $\langle {\bar q} q\rangle$. In QCD
  such a proportionality relation  is absent, the quark mass and the quark condensate in the chiral limit  being independent parameters.
   This feature of the soft-wall model comes from the choice of the terms in the $X$ field in the action  (\ref{action-0}) or  (\ref{action}), and could be corrected adding potential terms
$V(|X|)$ to the action \cite{Pomarol-scalars,various-pot}. In the following analysis we do not explore such a  possibility, but  simply assume for  $v(y)$ the form
 \be
 v(y)=\frac{m_q}{c} \, y + \frac{\sigma}{c^3} \, y^3 \,\,\, , \label{v}
 \ee
 the same choice done, e.g.,  in \cite{lebed},
considering  separately the cases where one or both the chiral symmetry breaking parameters are different from zero.

Now we  proceed to determine the  functions  $w_L$ and $w_T$ which  can be obtained, according to the AdS/CFT prescription,  by  a functional derivation of the effective 5d action (\ref{effective-action}).
Before the calculation,  we express the Chern-Simons action  (\ref{S-CS-boundary}) in terms of the weak background electromagnetic field
\be
S_{CS+b}=48 \, k_{CS}  \, d^{ab} \, {\tilde F}^{\mu \nu}_{em} \int d^5x \, A^b_\nu \, \partial_z V^a_\mu \,\,\, ,
\label{CSnew}
\ee
with $d^{ab}={1 \over 2}Tr[Q\{T^a,T^b\}]$, $Q$  the electric charge matrix as before, and ${\tilde F}^{\mu \nu}_{em}$  the field strength  corresponding to the external photon.  The electric charge matrix  obeys the Gell-Mann Nishijima relation:  $Q=T^3+ \displaystyle{Y \over 2}$, with $Y$  the  hypercharge that can be expressed in terms of the generators of $U(N_f)$: $T^a$ for $SU(N_f)$ and $T^0$ proportional to the baryon number matrix $B=\displaystyle{1 \over 3} \mathbf{1}$ that generates $U(1)$.  With two light flavors the relation is $Y=\displaystyle{B \over 2}$, while for  three flavors it is $Y=\displaystyle{1 \over 2}(B+S)$ where
$ S=Diag(0 , \, 0 , \, -1 )$ is the strangeness matrix. In this case $Y$ is proportional to the generator $T^8$ of $SU(3)$:
$Y=\displaystyle{1 \over \sqrt{3} } T^8$, so that  $Q=T^3 +\displaystyle{1 \over \sqrt{3} } T^8$ and ${\tilde F}^{\mu \nu}_{em}={\tilde F}^{3,\mu \nu}+ \displaystyle{1 \over \sqrt{3} } {\tilde F}^{8,\mu \nu}$.
 Notice that for a soft $x^\mu$-independent  electromagnetic field
its dual vector field is also independent of the fifth coordinate $z$,
so $\widetilde{F}^{\mu\nu}_{em}$   is placed out of the 5d integral in Eq.(\ref{CSnew}).
 
The action (\ref{CSnew}) can be used  to derive the expressions of $w_{L}(Q^2)$ and $w_{T}(Q^2)$.
Analogously to the decomposition  in (\ref{decomposition}),
 the correlation function of a vector  and an axial-vector current  in the external electromagnetic background field can be  written in terms of the functions $w_{L}$ and $w_{T}$:
\bea
d^{ab} \langle J_\mu^{V} J_\nu^{A} \rangle_{\tilde F}&\equiv & i \int d^4 x \, e^{i q x} \langle T \{J_\mu^{Va} (x) J_\nu^{Ab}(0) \} \rangle_{\tilde F}\nn \\
&=&  d^{ab} {Q^2 \over 4 \pi^2} P_{\mu \alpha}^\perp \left[P_{\nu \beta}^\perp w_T(Q^2) + P_{\nu \beta}^\parallel w_L(Q^2) \right] \, {\tilde F}^{\alpha \beta} \,\,.
\label{JJ-F}
\eea
The two terms in this expression, the one proportional to $P_{\mu \alpha}^\perp P_{\nu \beta}^\perp$ and the other one proportional to $P_{\mu \alpha}^\perp P_{\nu \beta}^\parallel$,  can be obtained by functional derivation of the action (\ref{effective-action}):
\bea
d^{ab} (2\pi)^{-4} \delta^4(q_1+q_2) \langle J_\mu^{V} J_\nu^{A} \rangle_{\tilde F}^{\perp \perp}&=&
{\delta^2 S_{CS+b} \over \delta V_{\mu0}^{a \perp}(q_1) \, \delta A_{\nu0}^{b \perp}(q_2) }\nonumber \\
d^{ab} (2\pi)^{-4} \delta^4(q_1+q_2) \langle J_\mu^{V} J_\nu^{A} \rangle_{\tilde F}^{\perp \parallel}&=&
{\delta^2 S_{CS+b} \over \delta V_{\mu0}^{a \perp}(q_1) \, \delta A_{\nu0}^{b \parallel}(q_2) } \label{funct-der} \,\,,
\eea
and from the comparison of  (\ref{JJ-F}) with (\ref{funct-der}) one finds:
\bea
w_L(Q^2)&=&-{2 N_c \over Q^2}\int_0^{\infty} dyA_\parallel(Q^2,y)  \partial_y V(Q^2,y) \label{wL-ris}
\\
w_T(Q^2)&=&-{2 N_c \over Q^2}\int_0^{\infty} dy A_\perp(Q^2,y) \partial_y V(Q^2,y) \label{wT-ris}\,\,.
\eea
The coefficient $2 N_c$ has been obtained fixing  the factor $k_{CS}$ in the Chern-Simons action (\ref{chernsimons}) to the value $\displaystyle k_{CS}=-{N_c \over 96 \, \pi^2}$; this permits
to recover the leading terms in the QCD OPE Eq.(\ref{basic-rel}),  as discussed below.

To see whether the expressions obtained  from Eqs.(\ref{wL-ris}) and (\ref{wT-ris})  match the QCD results  of  the previous Section,   the equations  for $V$, $A_\parallel$ and  $A_\perp$ must be  analyzed and solved.

\section{Determination of the functions $w_L$ and $w_T$}\label{numerics}
In order to compute the functions $w_L(Q^2)$ and $w_T(Q^2)$ using Eqs.(\ref{wL-ris}) and (\ref{wT-ris})   we need to analyze and  solve  the equations of motion
(\ref{EOM-V}) -(\ref{EOM-phi-pi-2}) for $V$, $A_\parallel$ and  $A_\perp$.
Equation (\ref{EOM-V})  for $V(Q^2,y)$ can be exactly solved with the boundary conditions $V(Q^2,0)=1$   and $V(Q^2,\infty)=0$,  yielding
\be
V(Q^2,y)=\Gamma \left( 1+ {{ Q}^2 \over 4 c^2} \right) U\left({ { Q}^2 \over 4 c^2},0,y^2 \right) \,\,\, ,
\label{V-sol}
\ee
with  $U$  the Tricomi confluent hypergeometric function.
The calculation is more difficult for $A_\perp$ and $A_\parallel$ since Eqs.(\ref{EOM-A}) and (\ref{EOM-phi-pi-2})  involve   the chiral symmetry breaking function $v(y)$.
Adopting the expression in (\ref{v}),  we  discuss separately  the cases:
\begin{itemize}
\item[A)] $m_q=\sigma=0$
\item[B)] $m_q\neq 0$, $\sigma=0$
\item[C)] $m_q=0$, $\sigma\neq0$
\item[D)] $m_q\neq 0$, $\sigma\neq0$ \,\,\, .
\end{itemize}

\subsection{$m_q=\sigma=0$}
If  both the chiral symmetry breaking parameters $m_q$ and $\sigma$ vanish, the equations of motion
 (\ref{EOM-A})-(\ref{EOM-phi-pi-2}) can be solved and provide  the results  $A_\parallel(Q^2,z)=1$,  and $A(Q^2,z)=A_\perp (Q^2,z)=V(Q^2,z)$
since Eqs.(\ref{EOM-A}) and (\ref{EOM-V}) coincide for $v(y)=0$.
Therefore,  the expressions (\ref{wL-ris}) and (\ref{wT-ris}) for the structure functions $w_L(Q^2)$ and $w_T(Q^2)$   become
\bea
w_L(Q^2)&=&-{2 N_c \over Q^2}\int_0^{\infty} dy \, \partial_y V(Q^2,y) = {2 N_c \over Q^2}\label{wL-ris-mqzero}
\\
w_T(Q^2)&=&-{2 N_c \over Q^2}\int_0^{\infty} dy \, V(Q^2,y) \partial_y V(Q^2,y) = {N_c \over Q^2}\label{wT-ris-mqzero}
\eea
using  the boundary conditions for $V(Q^2,y)$ at $y=0$ and $y \to +\infty$.
Equations (\ref{wL-ris-mqzero}), (\ref{wT-ris-mqzero}) show that the QCD results in  Eqs.(\ref{basic-rel-0}),(\ref{basic-rel}) in the case of chiral symmetry restoration  are recovered  in the holographic  approach.

\subsection{$m_q \neq 0$, $\sigma=0$}

In this case,  Eqs.(\ref{EOM-A}) and (\ref{EOM-V}) coincide  replacing  ${\tilde Q}^2 \to  {\tilde Q}^2+  {\tilde M}^2$,  where $ \displaystyle {\tilde M}^2={m_q^2 \, g_5^2 \over c^2}$. Therefore,  the solution of (\ref{EOM-A}) satisfying the conditions $A_\perp( Q^2,0)=1$ and $A_\perp(Q^2,\infty)=0$  is
\be
A_\perp(Q^2,y)=\Gamma \left( 1+ {{\tilde Q}^2 +  {\tilde M}^2\over 4} \right) U\left({ {\tilde Q}^2+  {\tilde M}^2 \over 4},0,y^2 \right)
\label{Aperp-sol-mq-si-sigma-no}
\,\,.
\ee
Also Eqs.(\ref{EOM-phi-pi-1}-\ref{EOM-phi-pi-2}) can be solved and yield
\bea
\pi(Q^2,y)&=& {{\tilde Q}^2 \over {\tilde M}^2}\left[1-A_\parallel(Q^2,y)\right]+\pi(Q^2,0)
\label{pi-sol-mq-si-sigma-no}\\
A_\parallel(Q^2,y) &=& {{\tilde M}^2 \over {\tilde Q}^2+  {\tilde M}^2} \left[1-\pi(Q^2,0)\right]A_\perp(Q^2,y) +{{\tilde Q}^2+{\tilde M}^2 \pi(Q^2,0)\over {\tilde Q}^2+{\tilde M}^2}
\label{Apar-sol-mq-si-sigma-no} \,\,.
\eea
These results lead to a relation between $w_T$ and $w_L$:
\be
w_L(Q^2)={2N_c \over Q^2} +{{\tilde M}^2 \left[1-\pi(Q^2,0)\right] \over {\tilde Q}^2+{\tilde M}^2} \left(w_T(Q^2)-{2N_c \over Q^2}\right) \,\,. \label{wT-vs-wL-B}
\ee
A critical role is played by the boundary condition $\pi(Q^2,0)$ of the chiral field solution of   (\ref{EOM-phi-pi-1}),(\ref{EOM-phi-pi-2}), an issue that  we shall examine  later on.
\subsection{$m_q = 0$, $\sigma \neq 0$}\label{sec:case3}

In this limit, the chiral limit, it is possible to determine the large $Q^2$ behavior of $w_L$ and $w_T$ by  the Green's function method described in  Appendix \ref{appB}.
An important point,  demonstrated  in the same appendix,  is that  $A_\parallel(Q^2,y)=1$ to all orders in the $1/Q^2$ expansion,  and at the same conclusion one  arrives considering the regularity of the solutions of the equation of motion, as discussed in Appendix \ref{appA}.
 The consequence, using (\ref{wL-ris}), is that
\begin{equation}
 w_L(Q^2) = \displaystyle\frac{2N_c}{Q^2} \,\,\, .
\end{equation}
Regarding $A_\perp$, the first correction appears at $\displaystyle {\CMcal O}\left(\frac{1}{Q^6}\right)$, and the resulting   modification in $w_T$ is
\begin{equation}
 w_T(Q^2)= \displaystyle\frac{N_c}{Q^2} - \tau \, g_5^2 \, \sigma^2 \, \displaystyle\frac{2N_c}{Q^8} + {\CMcal O}\left({1\over Q^{10}}\right)\;, \label{wT-mq}
\end{equation}
with $\tau=2.74286$ a numerical constant obtained  in the Appendix \ref{appBchiral} by the Green's function method.
The result in (\ref{wT-mq})  does not reproduce the  QCD one, Eq.(\ref{eq.wT-OPE-mq0}), in which the first power correction shows up  at $\displaystyle {\CMcal O}\left(\frac{1}{Q^6}\right)$ and is proportional to the
magnetic susceptibility $\chi$ of the quark condensate.

A comment concerning this discrepancy is in order.
The asymptotic conformal symmetry of QCD in the  Euclidean large $Q^2$ region suggests that AdS/CFT related methods can be used to describe strong interactions in this range of squared momenta. However,  QCD is  weakly coupled in this regime while, in principle, 
the gauge/gravity correspondence  relates a supergravity theory to a gauge field theory  which is  strongly coupled at  all scales.  Standing the conjecture,
the smallness of the QCD coupling at $Q^2\to \infty$ could enhance  the  stringy effects in the gravity dual. In bottom-up models,  this might imply a  mismatch of the ${\CMcal O}(\alpha_s)$ corrections,  as reported here in the case of the anomaly, or  a mismatch of condensate terms in the $1/Q^2$ expansion of different  correlation functions \cite{Colangelo:2007if}. A justification of the application of the holographic correspondence  in a regime different from the one in which it is expected to hold can be  found  observing that a few results computed in QCD through  expansions, including the $1/Q^2$ one, can be reproduced in dual models with various gravity backgrounds, as obtained in \cite{Erdmenger:2011sz}.

\subsection{$m_q\neq 0$, $\sigma\neq0$}
In this more general case, results can be obtained by the Green's function method in the large $Q^2$ limit, as we discuss afterward. There is also the possibility to work out analytical results in an
 interesting situation, in which $m_q^2$ and $\sigma^2$ terms in  the function $v(y)^2$ are neglected,  and only the term proportional to $m_q \sigma$ is considered.  In this case, the inclusion of $m_q^2$ terms can be done
subsequently in a straightforward way in the case of $A_\perp$.  Notice that, on dimensional ground, $\sigma^2$ terms will be suppressed by higher inverse powers of $Q^2$ and give  subleading
 contributions in a $1/Q^2$ expansion to be matched with the OPE in QCD.
 Therefore, we first concentrate  on the discussion of these analytic results.

Let us consider Eq.(\ref{EOM-A}) for  $A=A_\perp$. For $v^2(y)=\displaystyle{2 m_q \sigma \over   c^4} y^4$, defining the dimensionless parameter $\lambda= \displaystyle{2 g_5^2 m_q \sigma \over   c^4}$,
Eq.(\ref{EOM-A}) becomes:
\be
\partial_y^2 A-\left(2y+{1 \over y} \right) \partial_y A -{\tilde Q^2}\, A - \lambda \, y^2 A=0 \,\,\, , \label{eq-Aperp-mqsigma}
\ee
and its  solution satisfying the boundary  conditions $A(Q^2,0)=1$ and regularity for $y \to \infty$,   is
\be
A(Q^2,y)= e^{{y^2 \over 2} \left(1-\sqrt{1+\lambda} \right)}  \Gamma \left( 1+ { {\tilde Q^2} \over 4 \sqrt{1+\lambda}} \right) U\left( { {\tilde Q^2} \over 4 \sqrt{1+\lambda}}, 0, \sqrt{1+\lambda} \,y^2 \right)
\,\,.
\label{ris-esatto-Aperp-mqsigma}
\ee
The expansion of this solution at first order in $\lambda$ involves the function   $V(Q^2,y)$  in (\ref{V-sol}) and  its derivatives:
\be
A(Q^2,y)= V(Q^2,y)+{\lambda \over 2} \left[-\left(1+{y^2 \over 2} \right)V(Q^2,y)+{y \over 2} \,\partial_y V(Q^2,y)+\left({ {\tilde Q^2} \over 4}\right)^2 W \left({{\tilde Q^2} \over 4},0,y^2 \right) \right] \,\,\, ,
\label{Aperp-ris}
\ee
together with  the function $W$,  defined as
\be
W(a,b,c)=-\partial_a \, \left\{\Gamma(a) U(a,b,c)\right\}=\int_0^\infty dt  \, e^{-ct} t^{a-1}(1+t)^{b-a-1}\log \left({1+t \over t} \right)\,\,\, . \label{Wgeneric}
\ee
This function $W$ can  be related to $V$:
\be
W\left({{\tilde Q^2} \over 4},0,y^2 \right)=\left({ 4 \over {\tilde Q^2} }\right)^2  \left[V(Q^2,y)-{\tilde Q^2}\partial_{\tilde Q^2} V(Q^2,y) \right] \label{WversusV}
\ee
and satisfies the condition $W\left({{\tilde Q^2} \over 4},0,0 \right)=\left({ 4 \over {\tilde Q^2} }\right)^2$. The solution permits to compute the function $w_T$ at the first order in $\lambda$,
\be
w_T (Q^2)={N_c \over Q^2} \left\{ 1+{\lambda \over 2} \left(  (3 - {\tilde Q}^2) I_1 -1+ 2 {\tilde Q}^2 I_Q \right) \right\}
\ee
where
\bea
I_1&=&\int_0^\infty dy \,y^2 \, V(Q^2,y) \partial_y V(Q^2,y) =-{1 \over  \left[\Gamma\left( {{\tilde Q}^2 \over 4}\right)\right]^2 }G_{33}^{33}
\left( 1 \Bigg| \begin{array}{c} 1,1,3-\frac{{\tilde Q}^2}{4} \\ 2,3,1+\frac{{\tilde Q}^2}{4} \end{array} \right) \,\,\, , \nonumber \\
I_Q&=& \int_0^\infty dy \, \left( \partial_{\tilde Q^2} V\right)\,\partial_y V(Q^2,y)\,\,\, .
\eea
 $G_{33}^{33}$ is the Meijer's G function.  Expanding $I_1$ and $I_Q$  in inverse powers of $ {\tilde Q}^2$,
\bea
I_1&=&-{2 \over 3 \, {\tilde Q}^2} -{8 \over 5 \,  {\tilde Q}^4} -{352 \over 105 \, {\tilde Q}^6} +{\cal O}\left( {1 \over {\tilde Q}^8} \right) \nonumber \\
I_Q&=&{1 \over 6 \, {\tilde Q}^2} +{1 \over 5 \, {\tilde Q}^4} -{8 \over 105 \, {\tilde Q}^6} +{\cal O}\left( {1 \over {\tilde Q}^8} \right)\,\,,
\eea
gives the result, at ${\cal O}\left( {1 \over {\tilde Q}^6}\right)$,
\be
w_T(Q^2) ={N_c \over Q^2} \left( 1-{4 \,  \lambda \over 5 \, \tilde Q^4} \right)  \,\,\, . \label{wt-1}
\ee
Let us  discuss  the inclusion of ${\cal O}(m_q^2)$ terms.
The solution for $A(Q^2,y)$ can be obtained in a straightforward way solving Eq.(\ref{eq-Aperp-mqsigma}) after replacing $ {\tilde Q}^2 \to {\tilde Q}^2  + {\tilde M}^2$ where again $ {\tilde M}^2={m_q^2 g_5^2 \over c^2}$.
Hence, the solution is provided by Eq.(\ref{ris-esatto-Aperp-mqsigma}) (to all orders in $\lambda$) or by Eq.(\ref{Aperp-ris}) (at ${\cal O}(\lambda)$) performing such a replacement.
Neglecting terms of ${\cal O}(\lambda\, {\tilde M}^2)$,  $w_T$ gets a correction  which reads (up to  ${\cal O}( {\tilde M}^4)$)
\be
w_T^{({\tilde M}^2)}(Q^2)=-{2 N_c \over Q^2}\left( {\tilde M}^2\,I_Q +{ {\tilde M}^4 \over 2} \partial_{{\tilde Q}^2} I_Q \right)\,\,\, , \label{wT-M2}
\ee
so that, expanding in the inverse powers of ${\tilde Q}^2$, we find:
\be
w_T(Q^2)={ N_c \over Q^2}\left(1-{g_5^2 m_q^2 \over 3 Q^2}-{2g_5^2 m_q^2 c^2 \over 5 Q^4}+
{g_5^4 m_q^4 \over 6 Q^4}-{8g_5^2 m_q \sigma \over 5 Q^4}  \right) + {\cal O}\left({1 \over Q^8}\right) \,.
\label{wT-final}
\ee

Now we turn to the determination of  $A_\parallel$ and $w_L$ for $g_5^2 v^2=\lambda y^4$.
Using Eqs.(\ref{EOM-phi-pi-1}) and (\ref{EOM-phi-pi-2}), together with  the relation between $\tilde \phi$ and $A_\parallel$,  we obtain for  the function $f(Q^2,y)=\partial_y A_\parallel(Q^2,y)$ the equation
\be
\partial_y \left({1 \over y^2} \partial_y f \right)-\partial_y \left( \left( {2 \over y}+ {1 \over y^3} \right) f \right)-{{\tilde Q}^2 \over y^2} f -\lambda f=0 \,\,,
\label{eq-der-apar}
\ee
the regular solution of which is
\be
f( Q^2, y)=C_1 \, e^{{y^2 \over 2}(1-\sqrt{1+\lambda})}\, y \, U \left({ {\tilde Q^2} \over 4 \sqrt{1+\lambda}}, 0, \sqrt{1+\lambda} \,y^2 \right) =C_1 { y \over \Gamma \left(1+{ {\tilde Q^2} \over 4 \sqrt{1+\lambda}} \right)}\, A(Q^2,y)\,\,.
\label{ris-esatto-Apar-mqsigma}
\ee
The last equality comes from the comparison with (\ref{ris-esatto-Aperp-mqsigma}).

The integration constant $C_1$ is critical.
If  $C_1$ does not depend on $\lambda$, the solution in (\ref{ris-esatto-Apar-mqsigma}) is compatible with the condition $A_\parallel(Q^2,y)=1$ for  $\lambda\to 0$ only for $C_1=0$, with the consequence: $\displaystyle w_L={2N_c \over Q^2}$.
If $C_1$ depends on $\lambda$, it should vanish for $\lambda \to 0$ in order to fulfill that  condition for $A_\parallel$.  Assuming $C_1 \propto \lambda$ and  expanding  $f=f_0+\frac{\lambda}{2} f_1$, we obtain:

\bea
f_0(Q^2,y)&=&0 \nonumber \\
f_1(Q^2,y)&=&\tilde C_1 \,  y \, V( Q^2,y) \,\,\,
\eea
where ${\tilde C}_1$ does no more depend on $\lambda$. Hence $A_\parallel$ reads
\be
A_\parallel(Q^2,y)= 1+ \tilde C_1 \frac{\lambda}{2} \frac{1}{4-\tilde Q^2} \left[ 2 (y^2+1) V( Q^2,y)-y (\partial_y V( Q^2,y)) -2   \right]\,\,\, ,
\ee
and $w_L$ is given  by
\bea
w_L(Q^2) &=&{2N_c \over Q^2} \left( 1-{\lambda \over 2} \tilde C_1 I_1 \right) \,\, = \,\,  {2N_c \over Q^2} \left( 1+{\lambda \over 2} \tilde C_1 \big( \frac{2}{3 \tilde Q^2}+\frac{8}{5 \tilde Q^4} +\dots  \big) \right) \,\,\,
\label{wLgeneric}
\eea
in terms of $\tilde C_1$ which typically  is a function of ${\tilde Q}^2$.

In the general case $m_q\neq 0$, $\sigma\neq0$ analytical results are difficult to work out, and we rely, in the large $Q^2$ limit, on  the findings of the  Green's function  method in the Appendix \ref{appB}.
For  $w_T$  the result  of such a method reproduces  Eq.(\ref{wT-final}).
The result for $w_L$   can be expressed in terms of the  boundary condition of the chiral field $\pi(Q^2,0)$:
\begin{equation}\label{wLcompl}
 w_L(Q^2) = \frac{2N_c}{Q^2} - \left[1 - \pi(Q^2,0)\right]\,N_c\left[\frac{g_5^2\, m_q^2}{Q^4} +\frac{4g_5^2\, m_q \, \sigma}{Q^6} - \frac{2g_5^4\,m_q^4}{3Q^6} + {\CMcal O}\left(\frac{1}{Q^8}\right)\right]\,.
\end{equation}
Notice  that,  for $\sigma=0$, the results  (\ref{wT-final}),(\ref{wLcompl})
satisfy the  relation (\ref{wT-vs-wL-B}).

Considering Eqs.(\ref{wLgeneric}) and (\ref{wLcompl}), we conclude that, in the holographic model, the survival of quark mass  corrections to $w_L$ depends on integration constants:
they appear if  $\tilde C_1\neq 0$, or $\pi(Q^2,0)\neq 1$.
 Regardless of this, the relation (\ref{wLT-2})   between the functions $w_L$ and $w_T$ at large $Q^2$ is violated.

\section{$\Pi_{VV}-\Pi_{AA}$ in the soft-wall model}\label{sec:PiAAVV}

In \cite{Son:2010vc} the idea  has  been put forward that, in massless QCD and for any positive and negative $Q^2$, a relation should hold between  the structure function $w_T$ and the
left-right   two-point correlation  function,  defined by the difference
$\Pi_{LR}= \Pi^{VV}_\perp-\Pi^{AA}_\perp$ of the transverse invariant functions appearing in  the vector and axial-vector two-point correlators:
\begin{eqnarray}
\Pi_{\mu\nu}^{ab}(q)\,&=&   \,\,i\, \int d^4x\,\,e^{i q x}\,\,\langle0| T\{\, J^a_\mu (x) \, J^b_\nu(0)\}|0\rangle  \nonumber\\
&=&\, (q_\mu q_\nu -q^2 g_{\mu\nu})\, \delta^{ab} \, \Pi_\perp  (q^2)
\,+\, q_\mu q_\nu \, \delta^{ab} \, \Pi_\parallel(q^2)   \,\,\, , \label{PI_LR}
\end{eqnarray}
with vector   $J^{a}_\mu=\bar{q}\gamma_\mu T^a q$
and axial-vector currents   $J^{5a}_\mu=\bar{q}\gamma_\mu \gamma_5 T^a q$. The proposed relation reads
\be
w_T(Q^2)=\frac{N_c}{Q^2}+\frac{N_c}{F_\pi^2}Ê\, \Pi_{LR}(Q^2) \,\,\,\, , \label{son-yam}
\ee
with $F_\pi$ the pion decay constant.

Before commenting on the relation (\ref{son-yam}), let us focus on $\Pi_{LR}$ in our holographic approach;  it is worth reminding  that,
for  $m_q=0$,  $\Pi_{LR}$  is  an order parameter of the spontaneous chiral symmetry breaking, therefore it  represents an important  quantity for studying  the chiral structure of the theory.

In the  AdS/QCD soft-wall model the expression of $\Pi_{LR}(Q^2)$ requires the bulk-to-boundary propagators  $V(Q^2,y)$ and $A_\perp(Q^2,y)$ close to the UV brane $y=c \, z \to0$:
\begin{equation}
\Pi_{LR}(Q^2)= - \displaystyle\frac{ \, e^{-y^2}}{k_{YM}\, g_5^2\, \tilde{Q}^2}
\biggl(V(Q^2,y)\frac{\partial_y V(Q^2,y)}{y}\, -\, A_\perp(Q^2,y)\frac{\partial_y A_\perp(Q^2,y)}{y}   \biggr)\biggl|_{y\to0} \,.  \label{PiLR-SW}
\end{equation}
An  expression for $\Pi_{LR}(Q^2)$ can be  obtained  solving  the  equations of motion (\ref{EOM-V}) and (\ref{EOM-A})  for $V$  and  $A_\perp$   through a perturbative
expansion  in $\displaystyle \frac{1}{\tilde{Q}^2}$ using  the Green's function method, and the  details  of the computation can be found in appendix \ref{appBchiral}.
The large $\tilde Q^2$ expansion reads
\begin{eqnarray}\label{PiLR}
\Pi_{LR}(Q^2) & = & \,-\, \displaystyle\frac{1}{k_{YM}\, g_5^2}
\sum_{k=0}^\infty\displaystyle\frac{\zeta_k}{(\tilde{Q}^2)^k} \,.  \label{PiLR1}
\end{eqnarray}
As shown in Appendix \ref{appBchiral}, for $m_q=0$ the first nonvanishing coefficient in (\ref{PiLR1}) is
\begin{equation}
 \zeta_3 = \frac{8g_5^2\, \sigma^2}{5c^6}\, ,
\end{equation}
yielding
\begin{equation}
 \Pi_{LR}(Q^2)\,  = \, -\, \frac{N_c\, \sigma^2}{10\pi^2\, Q^6} + {\CMcal O}\left(\frac{1}{Q^8}\right)\, . \label{PILR-res}
\end{equation}
Therefore, the first term in the expansion of  $\Pi_{LR}$ is of  $\displaystyle {\CMcal O}\left(1/Q^{6}\right)$, with the same negative sign
found in QCD for the corresponding  dimension six condensate~\cite{SVZ,PI-Narison,PI-Narison-average}.
The result (\ref{PILR-res}) is  quite robust, since  additional contributions to $v=\sigma y^3/c^3$ with higher orders in $y$ would modify  $\Pi_{LR}$ at ${\CMcal O}(1/Q^{8})$ or beyond.

Concerning the relation (\ref{son-yam}),   at large $Q^2$ the difference between $V$ and $A$  is of ${\CMcal O}(1/{Q}^{6})$,
and this leads, for  $m_q=0$, to the  result obtained in  Sec.\ref{sec:case3} that the  leading power correction to $w_T$  is
$\displaystyle w_T(Q^2)=\frac{N_c}{Q^2}  \, \left(1\, + {\CMcal O}\big(\frac{1}{Q^{6}}\big)\right)$.  Considering that $\Pi_{LR}$ is given by  Eq.(\ref{PILR-res}),  we conclude that  the $Q^2$  dependences  of the two sides of the proposed equality (\ref{son-yam}) do not match, therefore  the validity of the relation (\ref{son-yam}) between $w_T$ and $\Pi_{LR}$ is not corroborated. A similar  result  has been found in the so-called
hard-wall model \cite{Son:2010vc}.

\section{Phenomenology  for $m_q=0$}\label{sec:PiLR}

For   $m_q=0$ and $Q^2\to 0$,  simple analytical results  for  $\Pi_{LR}$ and $w_T$ can be worked out.
In this case $g_5 \, v(z) = \Sigma y^3$
(with  $\Sigma = g_5 \sigma/c^3$),    therefore the regular solution  $A(0,y)=A_\perp(0,y)$ of Eq.(\ref{EOM-A}) can be written  in terms of the Airy function ${\rm Ai}(x)$:
\begin{eqnarray}
A(0,y) &=&\, e^{\frac{y^2}{2}} \,
\frac{\mbox{Ai}\left( \frac{{\Sigma}^2 y^2+1}{2^\frac{2}{3} {\Sigma}^\frac{4}{3}}
\right)}
{\mbox{Ai}\left( \frac{1}{2^\frac{2}{3} {\Sigma}^\frac{4}{3}} \right)  }\,\,\,\, .
\label{eq.AQ0}
\end{eqnarray}
The pion decay  constant is then provided by the relation \cite{Erlich:2005qh}
\begin{eqnarray}
F_\pi^2 &=& -\, \frac{1}{g_5^2 k_{YM}}\,\, c^2\,\, \frac{ \partial_y A(0,y)}{y}
\bigg|_{y\to 0}  \, \,\,=\,\,\,
 -\, \frac{N_c}{12\pi^2}\,\, c^2\,\, \frac{\partial_y A(0,y)}{y}
\bigg|_{y\to 0}  \, .\label{eq.Fpi-AdS}
\end{eqnarray}
The  function $w_T$   at $Q^2=0$  is related to a  chiral low-energy constant $C^W_{22}$, defined in   \cite{op6-lagr,Kampf:2005tz}:  they can be   both computed and read
\begin{eqnarray}
C^W_{22} &=& \frac{w_T(0)}{128\pi^2}
\,\,\,=\,\,\, -\, \frac{N_c}{64\pi^2 c^2}\, \int_0^\infty dy\,\,  A(0,y)\, f_V(y) \,\,\,\, ,
\label{eq.CW22-AdS}
\end{eqnarray}
with
\begin{equation}
f_V(y)  =  \frac{ \partial_y V(Q^2,y)}{ \tilde{Q}^2 } \bigg|_{\tilde{Q}^2\to 0}  =
-  \frac{y}{2}  \, e^{y^2} \, \Gamma(0,y^2)\,
\end{equation}
and $\Gamma(a,x)$ the incomplete gamma function.

Let us remark that,  as $F_\pi^2$ is of ${\CMcal O}(N_c)$, the derivative $\partial_y A(0,y)$ must be  ${\CMcal O}(N_c^0)$.
This requires that the parameter  $\sigma$ in the scalar background function $X_0(z)$    must be of  ${\CMcal O}(N_c^0)$ or smaller.
Indeed,  from the analysis of the  AdS/QCD effective action  including, together with  the background field $X_0(z)$, the dynamical scalar fields $S(x,z)$ \cite{Colangelo:2008us,Pomarol-scalars},  we  work out
the relation:   $\displaystyle \sigma=-\frac{8\pi^2}{N_c}\langle \bar{q}q\rangle$.
As a consequence, the numerical results for $F_\pi$ and $C^W_{22}$ from (\ref{eq.Fpi-AdS}) and (\ref{eq.CW22-AdS}), using the central value of the quark condensate  from QCD sum rules analyses
 $\langle \bar{q}q\rangle = - (0.24 \pm 0.01$~GeV$)^3$ (at the scale  $\mu=1$~GeV)
\cite{Colangelo:2000dp}, 
 together with  $c=M_\rho/2=0.388$~GeV and $N_c=3$, are
\begin{eqnarray}
F_\pi\, =\, 86.5\mbox{ MeV}  \, ,\qquad\qquad \qquad
C^W_{22}\,=\, 6.3 \times 10^{-3}\mbox{ GeV}^{-2}\, .
\end{eqnarray}
The experimental value of the neutral pion decay constant  is $F_\pi=92.2$~MeV. The low-energy constant  $C^{W}_{22}$ can be related to the slope at $Q^2=0$  of the $\pi^0 \to \gamma^* \gamma$ form factor:
$F(Q^2)=F(0)\left(1-\alpha \frac{Q^2}{M^2_{\pi^0}}\right)$.  The slope $\alpha$ has been measured: $\alpha=0.032\pm0.004$~\cite{PDG},  and the  relation with  $C_{22}^{W} $ has been obtained in the large $N_c$ limit:
$\displaystyle C_{22}^{W}= \frac{\alpha \, N_c}{64 \, \pi^2 M_{\pi^0}^2}$ ~\cite{Kampf:2005tz}.  The corresponding value is $C^{W}_{22}=(8.3\pm 1.3) \times 10^{-3}$~GeV$^{-2}$. Our result is also close to
an estimate by a resonance chiral theory, expressed in terms of the light vector meson mass $M_\rho$:
$\displaystyle C_{22}^{W}= \frac{N_c}{64\pi^2 M_\rho^2}=7.9 \times 10^{-3}$~GeV$^{-2}$
 ~\cite{anomaly-RChT}.

Finally, from  Eq.~(\ref{PILR-res})
it is also possible to obtain a determination of  the dimension six condensate in $\Pi_{LR}$, i.e. the coefficient of the $1/Q^6$ term in the $1/Q^2$  expansion,
\be
\mathcal{O}_6\, =\, -\, \frac{32\pi^2}{5 N_c}\,  \langle \bar{q}q\rangle^2
\,=\, -\, 4.0\,\times\, 10^{-3} \,\, \mbox{GeV}^6 \, ,
\ee
in reasonable agreement with QCD sum rule determinations~\cite{SVZ,PI-Narison,PI-Narison-average},
an average of which is provided in~\cite{PI-Narison-average}:
$\mathcal{O}_6=(-3.9\pm 0.8)\times 10^{-3}$~GeV$^6$.
Using different values of the parameters, namely the quark condensate  reported in \cite{Colangelo:2010et},  would not spoil the overall agreement
of the soft-wall results with the other determinations.

\section{Discussion and conclusions}\label{concl}

In the holographic approach with the Chern-Simons term in the action, the expressions (\ref{wL-ris}) and (\ref{wT-ris}) allow to determine $w_L$ and $w_T$ in terms of the functions $V$, $A_\parallel$ and $A_\perp$
which regulate  the vector and the axial-vector sectors in the dual model.
In the chiral $m_q=0$ limit,  the result  (\ref{basic-rel-0}) dictated  by the chiral anomaly is recovered for $w_L$. We have explicitely obtained such a result also in the case where the chiral condensate does not vanish,
looking at  regularity requirements for $A_\parallel$ (discussed in Appendix \ref{appA}), or calculating explicitly the large $Q^2$ expansion (in Appendix \ref{appB}). This  confirms that,  in the chiral limit,  $w_L$ is  essentially a topological quantity, it does not depend on the equations of motion but only on boundary conditions for $V$ and $A_\parallel$.  On the other hand, $w_T$
is dynamical  and  requires the solution of such  equations:   we have obtained that, when the chiral symmetry breaking field $v$ vanishes,  the result for $w_T$ reproduces the QCD expression and is related  to $w_L$  through  Eq.(\ref{basic-rel}).

Away from the chiral limit, the explicit solutions of the equations of motion for $V$, $A_\parallel$ and $A_\perp$ are needed to account for the quark mass corrections both in $w_L$ and $w_T$,  and for other  nonperturbative corrections to $w_T$. In the soft-wall model,  these  equations entail the  field $v$ which  breaks the chiral symmetry. We have chosen
a simple functional form  for $v(y)$, in which the quark mass term and the chiral condensate term are  specified,  Eq.(\ref{v}), in order to study separately the effect of these two quantities in $w_L$ and $w_T$, as well as in other observables and in a few low-energy constants, working out analytic solutions or expansions for
 large Euclidean squared momentum $Q^2$.  The effects of $v(y)$ in more involved models  in which this field dynamically arises, namely  by   appropriate potential terms in the 5d action, or in which the backreaction of matter on geometry is included, deserve  other  dedicated investigations.

Considering the correction induced by the  quark mass, we  recover in the $Q^2$ expansion  of  the structure function $w_T$ the next-to-leading  $\displaystyle {\CMcal O}\left(\frac{m^2}{Q^4}\right)$ term, see Eq.(\ref{wT-final}), but with an incorrect numerical factor
($-\frac{1}{4}$ instead of $+2$), and missing   the  $\displaystyle {\rm log}\left(\frac{m^2}{Q^2}\right)$ coefficient which appears in the corresponding one-loop QCD expression (\ref{wLT-1loop}). This is a consequence of the simplest inclusion of the quark mass in the holographic framework, and it is unlikely that it could be avoided without a radical modification of the ansatz  (\ref{v}).
The  $m_q\neq 0$ case also brings along a difficulty in fixing the value of the chiral field
$\pi(Q^2,y)$ at  the UV boundary $y=0$, which could not be established within our  $AV^*V$ analysis.
This boundary condition  affects  $A_\parallel$ too,  and therefore a possibility to fix the value of
$\pi(Q^2,0)$ (which is 1 in the chiral limit)  is   through  the $\Pi_{A_\parallel A_\parallel}$ correlation function at nonvanishing quark mass, a problem   requiring an  independent study. This boundary condition also influences  the relation (\ref{wT-vs-wL-B}) between $w_L$ and $w_T$.

For the general case in which both the quark mass and the quark condensate are different from zero, it is interesting to compare term by term
the subleading contributions  in the $1/Q^2$ expansion of $w_L$ and $w_T$ obtained  in QCD and in the holographic model.
Before doing that, let us remark that we have derived an exact analytical solution for $A_\perp(Q^2,y)$ in the case in which $v^2(y)$ can be approximated by the mixed $m_q \sigma y^4$ term,    Eq.(\ref{ris-esatto-Aperp-mqsigma}), obtaining also
that this analytical expression can be generalized  when the $m_q^2$ term is  included, by the substitution ${\tilde Q}^2 \to {\tilde Q}^2  + {\tilde M}^2$ in (\ref{ris-esatto-Aperp-mqsigma}). Such an achievement represents a step
towards a better understanding of the axial-vector sector in the soft-wall model. Moreover, it allows to obtain the structure function $w_T$ in the full range of squared momentum  $Q^2$ assuming this ansatz for $v$.

Considering the  expansion of $w_T$ for large Euclidean momenta $Q^2$, we have found a mismatch with the QCD result. Indeed, while in QCD, in the massless case, the next-to-leading contribution in $w_T$ is ${\CMcal O}(1/Q^6)$,
as in  Eq.(\ref{eq.wT-OPE-mq0}), we have found  a ${\CMcal O}(1/Q^8)$ term in Eq.(\ref{wT-mq}) in the dual model. Notice that in QCD the next-to-leading correction involves the magnetic susceptibility  $\chi$ of the quark condensate.
Analogously, in the massive case,  instead of finding a ${\CMcal O}(1/Q^4)$ term, which is also controlled by the susceptibility  $\chi$  in QCD, Eq.(\ref{wLT-2}), we have found a ${\CMcal O}(1/Q^6)$ correction, Eq.(\ref{wT-final}). Both  issues can be understood by  the perturbative Green's function expansion
in $1/\tilde{Q}^2$. Indeed,  for $m_q\neq 0$  the first $m_q \sigma$ correction
to $A_{\perp ,\parallel}$ shows up at next-to-next-to-leading order, i.e.  at ${\CMcal O}(1/Q^6)$ in $w_{T,L}$;  on the other hand, for $m_q=0$ the first correction from $v^2(y)$
is proportional to $\sigma^2$,  it appears at third order in perturbation theory,  hence at ${\CMcal O}(1/Q^8)$ in $w_T$.

A simple interpretation of this mismatch is that, in the soft-wall holographic model, the magnetic susceptibility of the chiral condensate turns out to vanish.
Going more deeply, the mismatch  implies that OPE terms in QCD involving operators like the tensor  $D=3$ operator $O_{\mu \nu}=\bar q\sigma_{\mu\nu} q$ and their matrix elements in the external electromagnetic field $F_{\mu \nu}$, have been missed in the dual approach, which instead produces an expansion similar to an OPE in vacuum.  A possible way out, which  deserves dedicated studies,  consists in explicitely including these
  $D=3$ operators  through  additional dual fields in the holographic model, a possibility  already considered in  different contexts \cite{Domokos:2011dn}.  Although the semiclassical limit of the theory in the AdS space is supposed to describe the nonperturbative regime of the gauge theory, it would be interesting to develop such new investigations in order to shed light, empirically, on the possibility of using the holographic approach in a regime which is not strongly coupled, as requested to compute the results of an OPE in QCD.

 The study of the left-right current correlator $\Pi_{LR}$ in the chiral limit has shown that other important features of QCD are reproduced  in the dual theory, namely the leading order of the $1/Q^2$  expansion and the value of the corresponding coefficient,
 which is in agreement with  the result found by traditional nonperturbative methods.
 Moreover, together with the value of the pion decay constant, also the low-energy parameter $C_{22}^W$, related to the slope at zero squared momentum transfer of the $\pi^0 \gamma^* \gamma$ form factor,  is close to the QCD value  and to the experimental measurement. On the other hand, corroboration of a proposed relation between $w_T$ and $\Pi_{LR}$, Eq.(\ref{son-yam}),  is not found.

To conclude, although we are not yet close to a formulation, in the bottom-up approach, of a holographic model in complete agreement with QCD, we have found  that, in spite  of its extreme simplicity and economicity, the soft-wall model reproduces more QCD properties that one could have expected.  Our study of the chiral $A V^* V$ anomalous vertex has shown several new features and difficulties, and has
 deepened  our understanding of the advantages and the limits of the model;  this represents a step towards  further  improvements. \\
 \noindent {\bf Note added.}
Another paper  discussing  the same   correlation function considered here   has recently appeared  \cite{Iatrakis:2011ht}.

\acknowledgments
We thank O. Domenech, K. Kampf, M. Knecht, A. Radyushkin and  N. Yamamoto for useful discussions. 
The work of J.J.S.C. has been partially supported by Universidad CEV Cardinal Herrera under contract PRCEU-UCH15/10.

\appendix

\section{Regular solutions for $\pi$ and $A_\parallel$ for $m_q=0$}\label{appA}

For $m_q=0$ there are constraints deriving from the requirement of regularity of $A_\parallel(Q^2,y)$ and $\pi(Q^2,y)$. Indeed,
in the gauge $A_z=0$,   the parallel component of the axial-vector field  and $\pi$ obey the  equations
\bea
e^\Phi \partial_y\Big(\frac{e^{-\Phi}}{y}  \partial_y A_\parallel \Big) \, +\,
\frac{g_5^2 v^2 }{y^3}    \, (\pi-A_\parallel)  &=&  0\, , \nn \\
\tilde Q^2 \,   \partial_y A_\parallel  \, +\, \frac{g_5^2 v^2 }{y^2}    \,  \partial_y \pi   &=&  0\, .   \label{parEOM}
\eea

For Euclidean momentum $Q^2>0$, one can define  the positive definite functional
\bea
f[A_\parallel,\pi]  &=& \int_\epsilon^\infty dy \,\, \frac{e^{-\Phi}}{y}
\Bigg\{  \tilde Q^2 (\partial_yA_\parallel)^2 +\frac{g_5^2 v^2}{y^2}   \Big[
\tilde Q^2 (\pi-A_\parallel)^2  + (\partial_y \pi)^2 \Big]
\Bigg\}\,\,\, \geq \,\,\, 0 \, .
\label{eq.functional}
\eea

If $A_\parallel$ and $\pi$ are solutions of the  equations of motion,
 the functional can be rewritten as a surface term:
\be
f[A_\parallel,\pi] \,\,\,=\,\,\, \int_\epsilon^\infty  dy\, \, \, \partial_y g(y) \,\,\,=\,\,\,
 g(\infty)-g(\epsilon)\,\,\,\geq \,\,\, 0 \, ,
\label{eq.functional2}
\ee
with
\be
g(y)\,\,=\,\,  - \,  \frac{e^{-\Phi}}{y}  \,  \tilde Q^2 \, (\pi-A_\parallel)\, \partial_yA_\parallel   \, .
\ee
Notice that $\partial_y g(y)$ vanishes  for values of $y$ where  $\partial_yA_\parallel=\pi- A_\parallel=0$.
On the other hand, $\partial_y g(y)$ is positive for the values of $y$ where both
$\partial_yA_\parallel\neq 0$ and $\pi\neq A_\parallel$.
Therefore, $g(y)$ is a monotonically growing function,
and   $g(\epsilon)< g(\infty)$   in correspondence to  nontrivial  solutions having
$\partial_y A_\parallel\neq 0$ and $\pi\neq A_\parallel$ in some range of  $y$.
If,  in addition, one assumes at most a power behavior $\sim y^n$ for the fields at  $y\to\infty$,
then $g(\infty)=0$ and one has $g(\epsilon)<0$ for nontrivial solutions.

Equation~(\ref{parEOM}) is a system of first order differential equations for the functions $\partial_y A_\parallel$ and $\pi-A_\parallel$,
and it has two  independent sets of solutions, which we label with the subscripts $(1)$ and $(2)$.
In the case $m_q=0$, $v(y) \stackrel{y\to 0}{=}{\CMcal O}( y^3)$ and
the analysis of the equations of motion  provides  the small $y$ behavior for  the two solutions,
\bea
\left(\ba{c} \partial_y A_{\parallel (1)} \\  \pi_{(1)}-A_{\parallel (1)} \ea\right)\,\,\,
\stackrel{y\to 0}{\sim}
\,\,\, \left(\ba{c} y^5 \\ y^0 \ea\right)\,, \qquad\qquad
\left(\ba{c}  \partial_y A_{\parallel (2)} \\\pi_{(2)}-A_{\parallel (2)}  \ea\right)\,\,\,
\stackrel{y\to 0}{\sim}
\,\,\,\left(\ba{c} y^1 \\  y^{-2} \ea\right)\,  . \label{cases}
\eea

If one assumes   $g(\infty)=0$, the functional~(\ref{eq.functional2}) becomes,   in correspondence to  the first solution:
\be
f[A_{\parallel(1)},\pi_{(1)}]\,\,\,=\,\,-\, g(\epsilon)_{(1)}\,\,\,
=\, \,\, {\CMcal O}(\epsilon^4)\,\,\,
\stackrel{\epsilon\to 0}{=}\,\,\, 0\, .
\ee
Therefore,  the first solution cannot be simultaneously nontrivial and regular
at $y\to \infty$  since, otherwise, $f[A_{\parallel(1)},\pi_{(1)}]$ would be different from zero.

On the other hand, the second solution (or a combination of the first and the second one)  makes not vanishing the functional~(\ref{eq.functional2}):
\be
f[A_{\parallel(2)},\pi_{(2)}]\,\,\,=\,\,-\, g(\epsilon)_{(2)}\,\,\,
=\, \,\, {\CMcal O}(\epsilon^{-2})\,\,\,
\stackrel{\epsilon\to 0}{\neq }\,\,\, 0\, ,
\ee
therefore it can be simultaneously  non-trivial and regular at $y\to \infty$.
However, from the second solution in (\ref{cases}) one finds that the combination $\pi_{(2)}-A_{\parallel(2)} \sim y^{-2}$ when $y\to 0$.
Since the ultraviolet boundary condition requires  $A_\parallel(\epsilon)=1$, the consequence is
that $\pi_{(2)}$ cannot be regular for   $y\to 0$.

The conclusion is that  the only possible solution,  regular both at small and large $y$,  is the trivial one,
\be
\pi(Q^2,y)\, -\,A_\parallel(Q^2,y) \,\,\,=\,\,\, \partial_yA_\parallel(Q^2,y) \,\,\,=\,\,\, 0\, ,
\ee
which   leads to $A_\parallel(Q^2,y)=\pi(Q^2,y)=1$ after imposing the ultraviolet boundary condition.

In the next appendix  we show  explicitly that the same conclusion
 follows from  the perturbative $1/\tilde{Q}^2$ expansion  in  the case $\Phi= y^2$ and $v=\sigma y^3/c^3$.

\section{Perturbative $1/\tilde Q^2$ expansion by the Green's function method}\label{appB}

The equations of motion (\ref{EOM-V})-(\ref{EOM-phi-pi-2}) can be solved perturbatively in $\beta=1/\tilde Q^2$ for large Euclidean $\tilde Q^2$ (small $\beta$) by defining the new variable $t=y\sqrt{\tilde Q^2}$.
In this  variable the equations read:
\begin{eqnarray}
 V''-\frac{1}{t}\,V'-V & = & 2\beta t\,V' \nonumber\\
 A_\perp''-\frac{1}{t}\,A_\perp'-A_\perp & = & 2\beta t\,A_\perp'   + (\beta \tilde M^2+ 2 \beta^2 \tilde M \Sigma t^2 +  \beta^3 \Sigma^2 t^4) A_\perp \\
 A_\parallel''-\frac{1}{t}\,A_\parallel'-A_\parallel & = & 2\beta t\,A_\parallel'+\left(\beta  \tilde M^2+2\beta^2 \tilde M\Sigma t^2+\beta^3\Sigma^2 t^4\right)\left(A_\parallel-\pi\right) \nonumber\\
 A_\parallel' & = & -\left(\beta \tilde M^2+2\beta^2 \tilde M\Sigma t^2+\beta^3\Sigma^2 t^4\right)\,\pi' \nonumber
\label{app:green1}
\end{eqnarray}
where $\displaystyle \tilde M=\frac{g_5 m_q}{c}$ and $\displaystyle \Sigma=\frac{g_5 \sigma}{c^3}$,  and the derivatives are with respect to $t$.  Expanding
\begin{eqnarray}
 V(Q^2,t)   =    \sum_{n=0}^\infty \beta^n V_n(t)\,,
&\qquad \qquad&
 A_\perp(Q^2,t)  = \sum_{n=0}^\infty \beta^n A_n^\perp(t)\,,\nonumber\\
 A_\parallel(Q^2,t)   =    \sum_{n=0}^\infty \beta^n A^\parallel_n(t)\,,
&\qquad \qquad&
 \pi(Q^2,t)  = \sum_{n=0}^\infty \beta^n \pi_n(t)\,,
\end{eqnarray}
we can solve the equations   order by order in $\beta$.
At  ${\CMcal O}(\beta^0)$ we have Bessel equations for $V$ and $A_\perp$:
\begin{eqnarray}
 V_0''-\frac{1}{t}\,V_0'-V_0 & = & 0 \nonumber\\
 {A^\perp_0}''-\frac{1}{t}\,{A^\perp_0}'-A^\perp_0 & = & 0
\end{eqnarray}
with boundary conditions $V_0(0)=A^\perp_0(0)=1$, therefore  the solution is  $V_0(t)=A^\perp_0(t)=t\,K_1(t)$, with  $K_1(t)$  the modified Bessel function of the second kind.
For the next orders, we consider separately the chiral limit, corresponding to  $\tilde M=0$,  and the case $\tilde M\neq0$.
The two cases have in common the feature that all the equations, to all orders $n$, are of the form
\begin{equation}
 f_n''-\frac{1}{t}\,f_n'-f_n={\CMcal F}\left[f_{n-1},f_{n-2},f_{n-3},\tilde M,\Sigma,t\right]
\end{equation}
where $f_n=V_n,A^\perp_n,A^\parallel_n,\pi_n$, and ${\CMcal F}$ is a functional that depends on the results found for the three previous orders (two for $n=2$, one for $n=1$)  and on the parameters.
This problem has a Green's function $G(t,s)$ which obeys the equation
\be\label{green-f}
\partial_t^2 G(t,s) - \frac{1}{t} \partial_t G(t,s) - G(t,s) = \delta(t-s)
\ee
and is given by
\begin{equation}\label{GF}
 G(t,s)=
\begin{cases}
 C_2(s)\,t\,I_1(t) & t<s\\
\\
 C_3(s)\,t\,K_1(t) & t>s
\end{cases}
\quad  {\rm where} \quad
\begin{cases}
 C_2(s) = -\displaystyle\frac{K_1(s)}{s\left[I_1(s)K_0(s)+I_0(s)K_1(s)\right]}\\
\\
 C_3(s) = -\displaystyle\frac{I_1(s)}{s\left[I_1(s)K_0(s)+I_0(s)K_1(s)\right]}
\end{cases}
\end{equation}
with $I_{0,1}$ and $K_0$ the modified Bessel functions of the first of the second kind, respectively.
The solutions  can be obtained  to all orders through (\ref{GF}).

\subsection{$\tilde M=0$, $\Sigma\neq0$}\label{appBchiral}

In this limit, the chiral limit, the first difference between the equations for $V$ and $A_\perp$ shows up at ${\CMcal O}(\beta^3)$ since  $V_i=A_i^\perp$ for $i=0,1,2$ while
$V_3\neq A_3^\perp$:
\begin{equation}
 A_3^\perp(t)=V_3(t)+\Sigma^2\,a_3(t)\,, \qquad {\rm with} \qquad a_3(t)=\int_0^\infty ds\,s^4\,G(t,s)\,V_0(s).
\end{equation}
These results, inserted in a large $Q^2$ expansion of (\ref{wL-ris}) and (\ref{wT-ris}), can be used to evaluate the ${\CMcal O}(1/Q^8)$ correction to $w_T$ reported in  Eq.~(\ref{wT-mq}),  coefficient of which is
\begin{equation}
 \tau = \int_0^\infty dt\,V_0'(t)\,a_3(t)=2.74286\,.
\end{equation}
The coefficients of Eq.~(\ref{PiLR}) can also be evaluated:
\begin{equation}
 \zeta_k  =  \displaystyle\lim_{t\to0}\displaystyle\frac{1}{t}\displaystyle\sum_{j=0}^k\left[V_{k-j}(t)V_j'(t) - {A^\perp}_{k-j}(t){A^\perp}_j'(t)\right]\,
\end{equation}
and, in particular,
\begin{equation}
 \zeta_3 = -\frac{g_5^2\, \sigma^2}{c^6}
  \lim_{t\to0} \frac{1}{t} \frac{d}{dt}\left[a_3(t) \, V_0(t)\right]=\frac{8g_5^2\, \sigma^2}{5c^6}\,.
\end{equation}

Concerning the longitudinal fields,  we have $A^\parallel_0(t)=\pi_0(t)=1$ identically, and $A^\parallel_n(t)=\pi_n(t)=0$, for all integers  $n\geqslant1$, since the equation for the $\pi_n$ fields
\begin{equation}
\pi_n''+\frac{3}{t}\pi'_n-\pi_n=0
\end{equation}
does not admit any solution which is regular in both the UV and the IR.
Therefore, we have $A_\parallel(t)=\pi(t)=1$   and,  as a consequence,   $\displaystyle w_L(Q^2)=\frac{2N_c}{Q^2}$  perturbatively to all orders  in $\beta=1/\tilde Q^2$.

\subsection{$\tilde M\neq0$, $\Sigma\neq0$}

In this  case  the first difference between the equations for $V$ and $A_\perp$ appears already at ${\CMcal O}(\beta)$:  $A_0^\perp=V_0$ and  $A_1^\perp(t)=V_1(t)+\tilde M^2\,\alpha_1(t)$, with
\begin{eqnarray}
 V_1(t) & = & 2\int_0^\infty ds\,G(t,s)\,s\,V_0'(s)\nonumber\\
 \alpha_1(t) & = & \int_0^\infty ds\,G(t,s)\,V_0(s)\,.
\end{eqnarray}
At ${\CMcal O}(\beta^2)$, we have
\begin{equation}
 A_2(t)=V_2(t)+\tilde M^2\,\beta_2(t)+\tilde M^4\,\gamma_2(t)+2 \tilde M \Sigma\,\delta_2(t)
\end{equation}
with
\begin{eqnarray}\label{perpgener2}
 V_2(t) & = & 2\int_0^\infty ds\,G(t,s)\,s\,V_1'(s)\nonumber\\
 \beta_2(t) & = & \int_0^\infty ds\,G(t,s)\,\left[V_1(s)+2s\,\alpha_1'(s)\right]\nonumber\\
 \gamma_2(t) & = & \int_0^\infty ds\,G(t,s)\,\alpha_1(s) \\
 \delta_2(t) & = & \int_0^\infty ds\,G(t,s)\,s^2\,V_0(s)\,.\nonumber
\end{eqnarray}
For the longitudinal fields, leaving the boundary condition for $\pi(Q^2,x)$ at $x=0$ unspecified, we have $A_0^\parallel(t)=1$ and
\begin{eqnarray}\label{parallgener2}
 A_1^\parallel(t) & = & \left[1-\pi_0(0)\right] \tilde M^2\left[V_0(t)-1\right]\nonumber\\
 {A_2^\parallel}'(t) & = & \left[1-\pi_0(0)\right]\left\{\tilde M^2\,V_1'(t)+2 \tilde M\Sigma\left[t^2\,V_0'(t)-V_1'(t)\right]+\tilde M^4\left[\alpha_1'(t)-V_0'(t)\right]\right\}\,.
\end{eqnarray}

Such expressions, inserted in a large $Q^2$ expansion of (\ref{wL-ris}) and (\ref{wT-ris}), allow to evaluate the functions $w_T$ and $w_L$ up to ${\CMcal O}(1/Q^6)$ by means of
 the integrals
\begin{eqnarray}
 \int_0^\infty dt\,V_0'(t)\,\delta_2(t)=\frac{2}{5}\,,
& \qquad \qquad &
 \int_0^\infty dt\,V_0'(t)\,\alpha_1(t)=\frac{1}{6}\,,\nonumber\\
 \int_0^\infty dt\,V_0'(t)\,\gamma_2(t)=-\frac{1}{12}\,,
& \qquad \qquad &
 \int_0^\infty dt\,V_1'(t)\,\alpha_1(t)=-\frac{1}{15}\,,\nonumber\\
 \int_0^\infty dt\,V_0'(t)\,\beta_2(t)=\frac{4}{15}\,,
& \qquad \qquad &
 \int_0^\infty dt\,V_0'(t)\,V_0(t)=-\frac{1}{2}\,,\nonumber\\
 \int_0^\infty dt\,V_0(t)\,V_1'(t)=\frac{1}{3}\,,
 & \qquad \qquad &
 \int_0^\infty dt\,t^2\,V_0'(t)\,V_0(t)=-\frac{2}{3}\,,\nonumber\\
 \int_0^\infty dt\,V_0(t)\,\alpha_1'(t)=-\frac{1}{6}\,. &&\,\nonumber
\end{eqnarray}



\begin{thebibliography}{99}

\bibitem{Maldacena:1997re}
  J.~M.~Maldacena,
  Adv.\ Theor.\ Math.\ Phys.\  {\bf 2}, 231 (1998)
  [Int.\ J.\ Theor.\ Phys.\  {\bf 38}, 1113 (1999)].
%
\bibitem{Witten:1998qj}
  E.~Witten,
  Adv.\ Theor.\ Math.\ Phys.\  {\bf 2}, 253 (1998).
%
\bibitem{Gubser:1998}
  S.~S.~Gubser, I.~R.~Klebanov and A.~M.~Polyakov,
  Phys.\ Lett.\  B {\bf 428},  105 (1998).

\bibitem{Witten2:1998}
  E.~Witten,
  Adv.\ Theor.\ Math.\ Phys.\  {\bf 2},  505 (1998).

\bibitem{altri}
  J.~Polchinski and M.~J.~Strassler,
  Phys.\ Rev.\ Lett.\  {\bf 88}, 031601 (2002);
%
D.~T.~Son and M.~A.~Stephanov,
Phys.\ Rev.\  D {\bf 69}, 065020 (2004);
%
  G.~F.~de Teramond, S.~J.~Brodsky,
 Phys.\ Rev.\ Lett.\  {\bf 94}, 201601 (2005) and
%
Phys.\ Rev.\ Lett.\  {\bf 102}, 081601 (2009);
%
  H.~Boschi-Filho, N.~R.~F.~Braga,
  JHEP {\bf 0305}, 009 (2003);
%
H.~Forkel, M.~Beyer, T.~Frederico,
JHEP {\bf 0707}, 077 (2007);
%
O.~Andreev, V.~I.~Zakharov,
Phys.\ Rev.\  D {\bf 74}, 025023 (2006);
%
  H.~R.~Grigoryan and A.~V.~Radyushkin,
  Phys.\ Lett.\  B {\bf 650}, 421 (2007) and
  Phys.\ Rev.\  D {\bf 76}, 095007 (2007);
%
  P.~Colangelo, F.~De Fazio, F.~Jugeau and S.~Nicotri,
  Phys.\ Lett.\  B {\bf 652}, 73 (2007);
  U.~Gursoy and E.~Kiritsis,
  JHEP {\bf 0802}, 032 (2008) and
  JHEP {\bf 0802}, 019 (2008);
%
  T.~Gherghetta, J.~I.~Kapusta, T.~M.~Kelley,
  Phys.\ Rev.\  {\bf D79},   076003 (2009);
  F.~Jugeau,
  Annals Phys.\  {\bf 325}, 1739 (2010).

  \bibitem{Erlich:2005qh}
J.~Erlich, E.~Katz, D.~T.~Son and M.~A.~Stephanov,
Phys.\ Rev.\ Lett.\  {\bf 95}, 261602 (2005).

\bibitem{Da Rold:2005zs}
  L.~Da Rold and A.~Pomarol,
  Nucl.\ Phys.\  B {\bf 721}, 79 (2005).

\bibitem{Karch:2006pv}
  A.~Karch, E.~Katz, D.~T.~Son and M.~A.~Stephanov,
  Phys.\ Rev.\  D {\bf 74}, 015005 (2006).

\bibitem{Colangelo:2008us}
  P.~Colangelo, F.~De Fazio, F.~Giannuzzi, F.~Jugeau and S.~Nicotri,
  Phys.\ Rev.\  D {\bf 78}, 055009 (2008).

\bibitem{top-down}
  J.~Erdmenger, N.~Evans, I.~Kirsch and E.~Threlfall,
  Eur.\ Phys.\ J.\  A {\bf 35}, 81 (2008).

\bibitem{Adler:1969gk}
  S.~L.~Adler,
  Phys.\ Rev.\  {\bf 177}, 2426 (1969).

\bibitem{Bell:1969ts}
  J.~S.~Bell and R.~Jackiw,
  Nuovo Cim.\  A {\bf 60}, 47 (1969).

\bibitem{Adler:1969er}
  S.~L.~Adler and W.~A.~Bardeen,
  Phys.\ Rev.\  {\bf 182}, 1517 (1969).

\bibitem{Vainshtein:2002nv}
  A.~Vainshtein,
  Phys.\ Lett.\  B {\bf 569}, 187 (2003).

\bibitem{Knecht:2003xy}
  M.~Knecht, S.~Peris, M.~Perrottet, E.~de Rafael,
  JHEP {\bf 0403}, 035 (2004).
  
  \bibitem{VVA-Vainshtein}
A.~Czarnecki, W.~J.~Marciano, A.~Vainshtein,
Phys.\ Rev.\  D {\bf 67} (2003) 073006 [Erratum-ibid.\  D {\bf 73} (2006) 119901].

\bibitem{Melnikov:2006qb}
  K.~Melnikov,
  Phys.\ Lett.\  {\bf B639}, 294 (2006).

\bibitem{Grigoryan:2008up}
H.~R.~Grigoryan and A.~V.~Radyushkin,
Phys.\ Rev.\  D {\bf 77}, 115024 (2008) and
%
Phys.\ Rev.\  D {\bf 78}, 115008 (2008).

\bibitem{Gorsky:2009ma}
  A.~Gorsky and A.~Krikun,
  Phys.\ Rev.\  D {\bf 79}, 086015 (2009).

\bibitem{Brodsky:2011xx}
  S.~J.~Brodsky, F.~-G.~Cao, G.~F.~de Teramond,
  arXiv:1105.3999 [hep-ph].

\bibitem{Hirn:2005nr}
J.~Hirn,  V.~Sanz,
JHEP {\bf 0512}, 030 (2005).

\bibitem{Son:2010vc}
D.~T.~Son and N.~Yamamoto,
arXiv:1010.0718 [hep-ph].

\bibitem{Knecht:2011wh}
M.~Knecht, S.~Peris, E.~de Rafael,
JHEP {\bf1110}, 048 (2011).
  
\bibitem{Knecht:2002hr}
  M.~Knecht, S.~Peris, M.~Perrottet, E.~De Rafael,
  JHEP {\bf 0211}, 003 (2002).

\bibitem{Hill:2006wu}
  C.~T.~Hill,
  Phys.\ Rev.\  D {\bf 73}, 126009 (2006).

\bibitem{Domokos:2007kt}
  S.~K.~Domokos, J.~A.~Harvey,
  Phys.\ Rev.\ Lett.\  {\bf 99},   141602 (2007).

\bibitem{Gorsky:2010xu}
  A.~Gorsky, P.~N.~Kopnin, A.~V.~Zayakin,
  Phys.\ Rev.\  {\bf D83},   014023 (2011)-
%

\bibitem{Kennedy:1999nn}
  C.~Kennedy, A.~Wilkins,
  Phys.\ Lett.\  {\bf B464},   206 (1999);
%
  P.~Kraus, F.~Larsen,
  Phys.\ Rev.\  {\bf D63},   106004 (2001);
%
  T.~Takayanagi, S.~Terashima, T.~Uesugi,
  JHEP {\bf 0103},   019 (2001).

\bibitem{Casero:2007ae}
  R.~Casero, E.~Kiritsis, A.~Paredes,
  Nucl.\ Phys.\  {\bf B787},   98 (2007);
  I.~Iatrakis, E.~Kiritsis, A.~Paredes,
  JHEP {\bf 1011},   123 (2010).
  
 \bibitem{Pomarol-scalars}
L.~Da Rold and A.~Pomarol,
JHEP {\bf 0601}, 157 (2006).

 \bibitem{various-pot}
B.~Batell and T.~Gherghetta,
 Phys.\ Rev.\ D {\bf 78}, 026002 (2008);
 T.~Gherghetta, J.~I.~Kapusta and T.~M.~Kelley,
 Phys.\ Rev.\  D {\bf 79}, 076003 (2009);
J.~I.~Kapusta and T.~Springer,
 Phys.\ Rev.\ D {\bf 81}, 086009 (2010);
 A.~Vega and I.~Schmidt,
 Phys.\ Rev.\  D {\bf 82}, 115023 (2010).
 
\bibitem{lebed}
  H.~J.~Kwee and R.~F.~Lebed,
  JHEP {\bf 0801}, 027 (2008).
  
  \bibitem{Colangelo:2007if}
 P.~Colangelo, F.~De Fazio, F.~Jugeau and S.~Nicotri,
 Int.\ J.\ Mod.\ Phys.\ A {\bf 24} (2009) 4177;
  H.~Forkel,
  PoS CONFINEMENT {\bf 8}, 184 (2008).
  
\bibitem{Erdmenger:2011sz}
  J.~Erdmenger, A.~Gorsky, P.~N.~Kopnin, A.~Krikun and A.~V.~Zayakin,
  JHEP {\bf 1103}, 044 (2011).
  
\bibitem{SVZ}
M.~A.~Shifman, A.~I.~Vainshtein and V.~I.~Zakharov,
Nucl.\ Phys.\  B {\bf 147}, 385 (1979);
Nucl.\ Phys.\  B {\bf 147}, 448 (1979).

 \bibitem{PI-Narison}
 J.~Bijnens, E.~Gamiz and J.~Prades,
 JHEP {\bf 0110}, 009 (2001);
 %
 S.~Friot, D.~Greynat and E.~de Rafael,
 JHEP {\bf 0410}, 043 (2004);
%
 V.~Cirigliano, E.~Golowich and K.~Maltman,
Phys.\ Rev.\  D {\bf 68}, 054013 (2003).

\bibitem{PI-Narison-average}
 S.~Narison,
 Phys.\ Lett.\  B {\bf 624}, 223 (2005).

\bibitem{op6-lagr}
  J.~Bijnens, L.~Girlanda and P.~Talavera,
  Eur.\ Phys.\ J.\  C {\bf 23}, 539 (2002);
%
  T.~Ebertshauser, H.~W.~Fearing and S.~Scherer,
  Phys.\ Rev.\  D {\bf 65}, 054033 (2002).

\bibitem{Kampf:2005tz}
  K.~Kampf, M.~Knecht and J.~Novotny,
  Eur.\ Phys.\ J.\  C {\bf 46}, 191 (2006).

\bibitem{Colangelo:2000dp}
P.~Colangelo, A.~Khodjamirian,
  In *Shifman, M. (ed.): At the frontier of particle physics, vol. 3* 1495-1576, World Scientific, Singapore
  [hep-ph/0010175].
  
 \bibitem{PDG}
K. Nakamura et al. (Particle Data Group),
J. Phys. G {\bf 37}, 075021 (2010) and 2011 partial update for the 2012 edition.

\bibitem{anomaly-RChT}
K.~Kampf and J.~Novotny,
arXiv:1104.3137 [hep-ph].

\bibitem{Colangelo:2010et}
  G.~Colangelo, S.~Durr, A.~Juttner, L.~Lellouch, H.~Leutwyler, V.~Lubicz, S.~Necco, C.~T.~Sachrajda {\it et al.},
  Eur.\ Phys.\ J.\  {\bf C71}, 1695 (2011).
  
\bibitem{Domokos:2011dn}
L.~Cappiello, O.~Cata, G.~D'Ambrosio,
  Phys.\ Rev.\  {\bf D82}, 095008 (2010);
O.~Cata,
 AIP Conf.\ Proc.\  {\bf 1317}, 328-333 (2011);
%
 S.~K.~Domokos, J.~A.~Harvey and A.~B.~Royston,
 JHEP {\bf 1105}, 107 (2011);
  R.~Alvares, C.~Hoyos and A.~Karch,
  arXiv:1108.1191 [hep-ph].

\bibitem{Iatrakis:2011ht}
  I.~Iatrakis, E.~Kiritsis,
  arXiv:1109.1282 [hep-ph].
  
\end{thebibliography}
\end{document}